\documentclass[%
preprint,
showpacs,
nofootinbib,
 amsmath,amssymb,
 aps,
prd,
]{revtex4-1}

\usepackage{graphicx}
\usepackage{dcolumn}
\usepackage{bm}

\newcommand{\nn}{\nonumber}
\newcommand{\ii}{{\hspace{0.7pt}\mathrm i\hspace{0.7pt}}}

\newcommand{\bk}{{\boldsymbol{k}}}
\newcommand{\bn}{{\boldsymbol{n}}}
\newcommand{\bx}{{\boldsymbol{x}}}

\newcommand{\bRR}{{\mathbb{R}}}
\newcommand{\bZZ}{{\mathbb{Z}}}

\newcommand{\cF}{{\mathcal F}}

\newcommand{\rmd}{{\mathrm d}}

\newcommand{\bra}[1]{\langle #1 \vert}
\newcommand{\ket}[1]{| #1 \rangle}

\newcommand{\abs}[1]{{\vert #1 \vert}}

\newcommand{\Tr}{\mathop{\mathrm{Tr}}\nolimits}
\newcommand{\ad}{\mathop{\mathrm{ad}}\nolimits}
\newcommand{\eq}{\rm{eq}}
\newcommand{\PP}{{\mathcal P}}
\newcommand{\QQ}{{\mathcal Q}}
\newcommand{\cH}{{\mathcal H}}

\newcommand{\sfx}{\mathsf{x}}
\newcommand{\sfy}{\mathsf{y}}
\newcommand{\dsfrac}[2]{\displaystyle\frac{#1}{#2}}
\newcommand{\afrac}[2]{\genfrac{}{}{0pt}{}{#1}{#2}}

\allowdisplaybreaks[3]

\begin{document}

\title[Master equation for the Unruh-DeWitt detector]{Master equation for the Unruh-DeWitt detector\\
and the universal relaxation time in de Sitter space}

\author{Masafumi Fukuma}
\email{fukuma@gauge.scphys.kyoto-u.ac.jp}
\affiliation{
 Department of Physics, Kyoto University  \\
Kyoto 606-8502, Japan
}%
\author{Yuho Sakatani}%
\email{yuho@cc.kyoto-su.ac.jp}
\affiliation{%
Maskawa Institute for Science and Culture,\\ Kyoto Sangyo University, 
Kyoto 603-8555, Japan
}%
\author{Sotaro Sugishita}%
\email{sotaro@gauge.scphys.kyoto-u.ac.jp}
\affiliation{
 Department of Physics, Kyoto University  \\
Kyoto 606-8502, Japan
}%


\date{\today}
\begin{abstract}

We derive the master equation 
that completely determines the time evolution of the density matrix 
of the Unruh-DeWitt detector in an arbitrary background geometry. 
We apply the equation 
to reveal a nonequilibrium thermodynamic character of de Sitter space. 
This generalizes an earlier study on the thermodynamic property 
of the Bunch-Davies vacuum 
that an Unruh-DeWitt detector staying in the Poincar\'e patch 
and interacting with a scalar field in the Bunch-Davies vacuum 
behaves as if it is in a thermal bath of finite temperature. 
In this paper, instead of the Bunch-Davies vacuum, 
we consider a class of initial states of scalar field, 
for which the detector behaves 
as if it is in a medium that is not in thermodynamic equilibrium 
and that undergoes a relaxation to the equilibrium 
corresponding to the Bunch-Davies vacuum. 
We give a prescription for calculating the relaxation times 
of the nonequilibrium processes. 
We particularly show that, 
when the initial state of the scalar field 
is the instantaneous ground state at a finite past, 
the relaxation time is always given by a universal value of 
half the curvature radius of de Sitter space. 
We expect that the relaxation time gives a nonequilibrium thermodynamic quantity 
intrinsic to de Sitter space.

\end{abstract}


\maketitle

\section{Introduction}
\label{sec:introduction}

The concept of particles is known to depend on observers. 
Even in the Poincar\'e-invariant Minkowski vacuum, 
an observer with constant acceleration sees a thermal particle spectrum 
\cite{Unruh:1976db}. 
The Unruh-DeWitt detector \cite{Unruh:1976db,dewitt} was introduced 
as a tool of thought experiment  
to give an intuitive understanding of such thermal character of spacetime 
(see also \cite{Birrell:1982ix,Takagi:1986kn} and references therein).  
This is a detector weakly interacting with a matter quantum field 
in a certain vacuum state, 
and one can study the thermal character of spacetime 
through the density distribution of the detector.

\begin{figure}[htbp]
\begin{center}
\includegraphics[width=6cm]{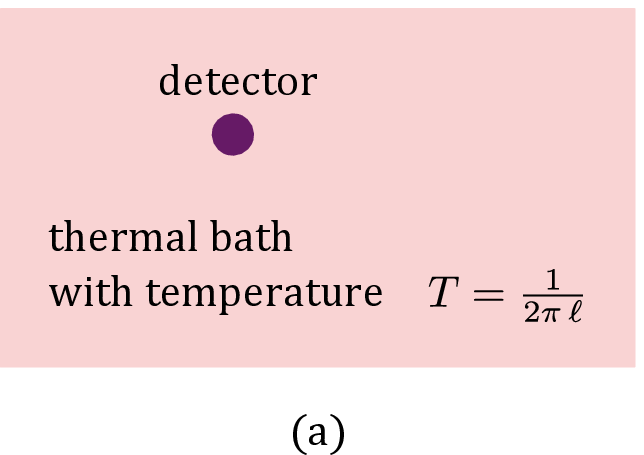}
\hspace{1cm}
\includegraphics[width=6cm]{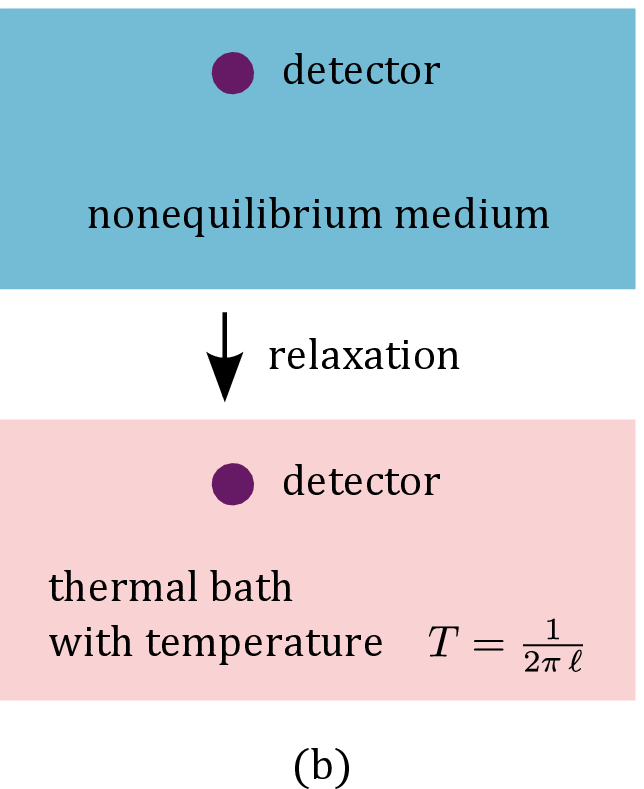}
\begin{quote}
\caption{The Unruh-DeWitt detector in the Poincar\'e patch. 
The scalar field is initially 
in (a) the Bunch-Davies vacuum 
or (b) a class of states deviated slightly from the Bunch-Davies vacuum.
\label{fig:medium}
}
\end{quote}
\end{center}
\vspace{-3ex}
\end{figure}

Various spacetimes have been examined with the Unruh-DeWitt detector, 
including de Sitter space. 
Although free scalar field in de Sitter space has a large family of de Sitter-invariant vacua 
(called the $\alpha$-vacua) \cite{Mottola:1984ar,Allen:1985ux},  
the Bunch-Davies vacuum (or the Euclidean vacuum) \cite{Bunch:1978yq} 
is regarded as the most natural vacuum 
because this satisfies the Hadamard condition. 
It is actually only the Bunch-Davies vacuum which exhibits a thermal property 
\cite{Birrell:1982ix,Spradlin:2001pw,Bousso:2001mw,Figari:1975km,Gibbons:1977mu};  
If one places a detector at $\bx={\mathbf0}$ 
in the Poincar\'e patch, 
\begin{align}
 \rmd s^2 = \ell^2\,\frac{{}-\rmd \eta^2+\rmd \bx^2}{\eta^2}
  \quad\bigl(-\infty<\eta<0\bigr)\,,
\end{align}
where scalar field is initially in the Bunch-Davies vacuum, 
then the density distribution of the detector evolves 
through the interaction with the scalar field 
and eventually reaches the Gibbs distribution of temperature $T=1/2\pi\ell$  
($\ell$ being the curvature radius of de Sitter space), 
irrespectively of the initial form of the density distribution  
[see Fig.~\ref{fig:medium} (a)]. 
This implies that the detector behaves 
as if it is in a thermal bath 
of temperature $T=1/2\pi\ell$ 
when it is placed in the Bunch-Davies vacuum.

The specialty of the Bunch-Davies vacuum may be understood as follows. 
We first notice that 
there is no global timelike Killing vector in the Poincar\'e patch  
and thus the Hamiltonian of scalar field has an explicit time dependence. 
This implies that one cannot define a time-independent ground state 
and can only introduce the instantaneous ground state $\ket{0_\eta}$ 
at each instant $\eta$\,. 
As is investigated in detail in \cite{Fukuma:2013mx}, 
the Bunch-Davies vacuum $\ket{\rm BD}$ 
can be characterized as the ground state at the infinite past, 
$\ket{\rm BD}=\lim_{\eta\to-\infty}\ket{0_\eta}$. 
Thus, one may regard the Bunch-Davies vacuum 
as a medium (surrounding the detector) 
which already undergoes a sufficiently long time evolution 
to reach a thermodynamic equilibrium state.

If instead the scalar field is initially in a certain class of states, 
the Unruh-DeWitt detector may behave 
as if it is surrounded by a medium 
that is not in thermodynamic equilibrium  
and that undergoes a relaxation to the equilibrium 
corresponding to the Bunch-Davies vacuum [see Fig.~\ref{fig:medium} (b)]. 
It should be possible to investigate the relaxation process of the surrounding medium 
by observing the time evolution of the density distribution of the detector. 
This analysis may give useful information 
on the nonequilibrium thermodynamic character 
intrinsic to de Sitter space.

The main purpose of this paper is to develop a machinery 
for describing such nonequilibrium dynamics 
and to calculate the relaxation times of the surrounding media. 
For this, we first develop a general framework 
to treat an Unruh-DeWitt detector in arbitrary background geometry 
and derive the master equation 
which completely determines a finite time evolution 
of the density matrix of the detector. 
We then apply this framework to a detector in de Sitter space. 
We show that if the initial state of the scalar field is chosen 
such that its Wightman function has the same short distance behavior 
as that of the Bunch-Davies vacuum, 
then the density distribution of any detector placed there 
exhibits a relaxation to the Gibbs distribution, 
with a relaxation time proportional to $\ell$ 
(measured in the proper time of the detector). 
In particular, if we take the initial state 
as the instantaneous ground state  
at a {\em finite} past (say, at $\eta_0$),
the relaxation time is always given by a universal value $\ell/2$.

In order to avoid possible confusions, 
we here stress that there can be two kinds of relaxation times. 
The first is the relaxation time that may exist 
even when the detector is placed in a thermal bath 
[see Fig.~\ref{fig:medium} (a)].  
This is the period of time 
it takes for the detector to reach the Gibbs distribution 
from a given initial density distribution. 
This kind of relaxation time can be neglected if one considers an ideal detector 
which can get adjusted to its environment instantaneously. 
Another kind of relaxation time, in which we are interested, 
is the period of time it takes for the nonequilibrium environment 
to reach a thermodynamic equilibrium state [see Fig.~\ref{fig:medium} (b)]. 
This relaxation time should not depend on details of the detector 
or on the form of interaction between the detector and the scalar field, 
and is related to the nonequilibrium dynamics 
intrinsic to de Sitter space.

This paper is organized as follows. 
In section \ref{sec:master}, 
adopting the method of projection operator \cite{kubo}, 
we first derive the master equation  
which describes the time evolution of the density matrix of the detector. 
Then, after justifying a Markovian approximation, 
we derive a simplified form of the master equation 
which enables us to study 
the relaxation behavior of the density distribution analytically. 
In section \ref{sec:dS}, 
we apply the framework to an Unruh-DeWitt detector 
in the Poincar\'e patch of de Sitter space. 
We consider a situation 
where the initial state of scalar field is chosen 
such that its Wightman function has the same short distance behavior 
as that of the Bunch-Davies vacuum. 
We compute the transition rate matrix of the density distribution 
of the detector, 
and show that the density distribution exhibits the expected relaxation 
to the equilibrium corresponding to the Bunch-Davies vacuum 
with the relaxation time of the form $\ell/\alpha$\,, 
where the constant $\alpha$ is determined 
by the asymptotic form of the change of the Wightman function 
from that of the Bunch-Davies vacuum. 
In section \ref{sec:fintie_past}, 
we consider a particular case 
where the initial state of scalar field 
is the instantaneous ground state $\ket{0_{\eta_0}}$ 
at a finite past $\eta=\eta_0$\,, 
and show that the relaxation time is always given by a universal value $\ell/2$\,, 
irrespectively of the value of $\eta_0$ or the form of interaction 
between the detector and the scalar field. 
Section \ref{sec:conclusion} is devoted to discussions and conclusion. 
We collect miscellaneous formulas in appendices.

\section{Master equation for the density matrix of an Unruh-DeWitt detector}
\label{sec:master}

\subsection{Setup}
\label{subsec:master_setup}

\begin{figure}[htbp]
\begin{center}
\includegraphics[width=4.5cm]{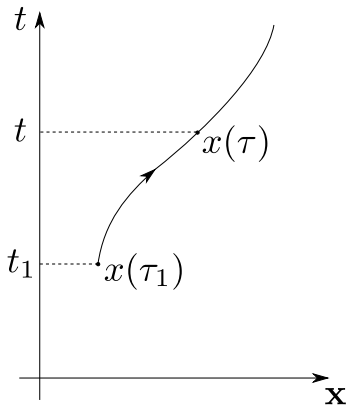}
\begin{quote}
\caption{The trajectory of an Unruh-DeWitt detector, 
which starts interacting with a scalar field $\phi(x)$ at time $t_1$\,. 
The time coordinate $t$ will be denoted by $\eta$ 
when it is the conformal time in the Poincar\'e patch. 
\label{fig:detector}
}
\end{quote}
\end{center}
\vspace{-3ex}
\end{figure}

We consider an Unruh-DeWitt detector 
in $d$-dimensional spacetime with background metric 
$\rmd s^2=g_{\mu\nu}(x)\,\rmd x^\mu\,\rmd x^\nu$ 
$(\mu,\nu=0,1,\ldots,d-1)$, 
the detector being interacting with a scalar field
$\phi(x)$ of mass $m$\,. 
We assume that the detector has a sufficiently large mass 
so that it can be treated as moving along a classical trajectory%
\footnote{
We also write the trajectory as $x^\mu(\tau)=\bigl(t,\bx(t)\bigr)$ 
using the functional relation $t=t(\tau)$ or $\tau=\tau(t)$ 
with $\rmd t/\rmd \tau >0$\,. 
\label{fn:trajectory}}
$x^\mu(\tau)=\bigl(t(\tau),\bx(\tau)\bigr)$, 
where $\tau$ is a proper time of the trajectory 
(see Fig.~\ref{fig:detector}). 

For a quantum mechanical description of the system, 
we introduce the Hilbert space $\cH^{\rm tot}$ 
which is the tensor product of those of the detector and the field,
\begin{align}
 \cH^{\rm tot}=\cH^{\rm d}\otimes\cH^{\phi}\,.
\end{align}
The total Hamiltonian then takes the following form in the Schr\"odinger picture: 
\begin{align}
 H^{\rm tot}(t) = H^{\rmd}\,
 \frac{\rmd \tau(t)}{\rmd t} \otimes 1
 + 1\otimes H^{\phi}(t)+V(t)\,.
\end{align}
Here, $H^{\rm d}$ is the Hamiltonian of the detector 
associated with its proper time, 
and we assume that $H^{\rm d}$ does not depend on $\tau$ (or on $t$), 
denoting its eigenstates and eigenvalues by $\ket{m}$ and $E_m$\,, 
respectively: 
$H^{\rmd}\ket{m}=E_m \ket{m}$\,.
$H^\phi(t)$ is the Hamiltonian of free scalar field 
and may depend on time explicitly 
through an explicit time dependence of the metric. 
$V(t)$ stands for the interaction between the detector and the field, 
and we assume that it suddenly starts at time $t_1$ 
in the form of monopole interaction,%
\footnote{
Recall that this is in the Schr\"odinger picture. 
We later shall take an average over the start-up time $t_1$ 
since one usually needs a certain period of time 
to specify the initial density distribution of the detector.}
\begin{align}
 V(t)=\lambda\,\mu\, \frac{\rmd \tau(t)}{\rmd t}
  \otimes\phi\bigl(\bx(t)\bigr)\,
  \theta(t-t_1)\,.
\end{align}
Here, $\lambda$ is a dimensionless coupling constant,
$\bx(t)$ represents the position of the detector at time $t$ 
(see footnote \ref{fn:trajectory}), 
and $\mu$ is an operator acting on $\cH^{\rm d}$\,, 
which we again assume to be time independent. 
The time evolution of the total density matrix $\rho^{\rm tot}(t)$ 
is given by 
\begin{align}
 \rho^{\rm tot}(t) 
  = U^{\rm tot}(t,t')\,\rho^{\rm tot}(t')\,\bigl[U^{\rm tot}(t,t')\bigr]^{-1}\,,
\end{align}
where $U^{\rm tot}(t,t')$ is the time evolution operator in the Schr\"odinger picture, 
\begin{align}
 U^{\rm tot}(t,t')\equiv
  {\rm T}\exp\Bigl(-\ii\int_{t'}^t\!\!\rmd t'\,H^{\rm tot}(t')\Bigr)\,.
\end{align}

We consider a situation where one can only measure 
observables associated with the detector, 
such as the matrix elements $\mu_{mn}=\bra{m} \mu \ket{n}$. 
Then the maximum information one can get from the system 
is the reduced density matrix $\rho(t)$ 
which is defined as the partial trace of the total density matrix $\rho^{\rm tot}(t)$ 
over $\cH^\phi$\,,
\begin{align}
 \rho(t) \equiv \Tr_{\phi}\rho^{\rm tot}(t)\,.
\end{align}
The time evolution of $\rho(t)$ should be uniquely determined 
once one specifies the form of interaction 
and the initial condition for $\rho^{\rm tot}(t)$\,. 
Since there had been no interaction 
between the detector and the field before time $t_1$\,, 
we may let the total density matrix take the following factorized form at $t_1$\,: 
\begin{align}
 \rho^{\rm tot}(t_1) \equiv \rho^{\rm d}(t_1) \otimes \rho^{\rm \phi}(t_1)
  =\rho(t_1) \otimes \rho^{\rm \phi}(t_1)\,. 
\label{init_rho}
\end{align}
The latter equality can be easily seen by noting that
$
 \rho(t_1)=\Tr_{\phi}\rho^{\rm tot}(t_1)=\rho^{\rm d}(t_1) 
$\,.

\subsection{Master equation}
\label{subsec:master_eqn}

The time evolution of $\rho(t)$ can be best analyzed  
if we go over to the interaction picture 
by decomposing the total Hamiltonian to 
\begin{align}
 H^{\rm tot}(t) &=H^{\rm tot}_0(t)+V(t)\,,
\\
 H^{\rm tot}_0(t) &\equiv H^{\rmd}\,\frac{\rmd\tau(t)}{\rmd t}\otimes 1
 + 1\otimes H^{\phi}(t)
\end{align}
and treat $V(t)$ as a perturbation.
The time evolution operator $U^{\rm tot}(t,t_1)$ 
is then decomposed to the unperturbed and perturbed parts as
\begin{align}
 U^{\rm tot}(t,t_1) = U^{\rm tot}_0(t,t_1)\,\,U^{\rm tot}_I(t,t_1)
\end{align}
with
\begin{align} 
 U^{\rm tot}_0(t,t_1) &\equiv  {\rm T}\exp\Bigl(
  -\ii\int^{t}_{t_1}\!\!\rmd t^{\prime}\, H^{\rm tot}_0(t^{\prime})\Bigr)
  \nn\\
  &= e^{{}-\ii H^{\rm d}\cdot\, (\tau-\tau_1)} \otimes
  {\rm T} e^{-\ii\int_{t_1}^t\rmd t' H^\phi(t')} \,,
\\
 U^{\rm tot}_I(t,t_1) &\equiv {\rm T}\exp\Bigl(-\ii\int^{t}_{t_1}\!\!\rmd t'\,
  V_I( t')\Bigr)\,.
\end{align}
Here, $V_I(t)$ is defined by 
\begin{align}
 V_I(t)&\equiv \bigl[U^{\rm tot}_0(t,t_1)\bigr]^{-1}
  V(t)\, U^{\rm tot}_0(t,t_1)
\nn\\
 &= \lambda\,\frac{\rmd \tau}{\rmd t}\,\mu_I(\tau) \otimes 
  \phi_I\bigl(x(\tau)\bigr)\,\theta(t-t_1)\,, 
\label{vint}
\end{align}
where $\mu_I(\tau)$ is given by 
\begin{align}
 \mu_I(\tau)
  &\equiv e^{\ii H^{\rm d}\cdot\,(\tau-\tau_1)}\,\mu\,e^{{}-\ii H^{\rm d}\cdot\,(\tau-\tau_1)}
  \nn\\
  &=\sum_{m,n}\,e^{\ii (E_m-E_n)(\tau-\tau_1)}\,\mu_{mn}\,
  \ket{m}\bra{n}\,,
\end{align}
and the operator 
\begin{align}
 \phi_I\bigl(x(\tau)\bigr)&\equiv \phi_I\bigl(t,\bx(t)\bigr)
 \nn\\
  &= \bigl[{\rm T} e^{-\ii\int_{t_1}^t\rmd t' H^\phi(t')} \bigr]^{-1}\,
  \phi\bigl(\bx(t)\bigr)\,{\rm T} e^{-\ii\int_{t_1}^t\rmd t' H^\phi(t')}
\end{align}
satisfies the free Klein-Gordon equation associated with the metric 
$\rmd s^2=g_{\mu\nu}(x)\,\rmd x^\mu\,\rmd x^\nu$\,. 
Accordingly, the density matrix in the interaction picture is given by 
\begin{align} 
 \rho^{\rm tot}_I(t)\equiv \bigl[U^{\rm tot}_0(t,t_1)\bigr]^{-1} 
  \rho^{\rm tot}(t)\, U^{\rm tot}_0(t,t_1)
  = U^{\rm tot}_I(t,t_1)\,\rho^{\rm tot}(t_1)\,\bigl[U^{\rm tot}_I(t,t_1)\bigr]^{-1}\,,
\end{align}
and satisfies the von Neumann equation of the form 
\begin{align}
 \frac{\rmd}{\rmd t}\rho^{\rm tot}_I(t)
  ={}-\ii\,[V_I(t),\rho^{\rm tot}_I(t)]
  \equiv -\ii\ad_{V_I(t)} \rho^{\rm tot}_I(t) \,.
\label{vn}
\end{align}

This von Neumann equation can be rewritten 
to an equation involving only $\rho_I(t)\equiv\Tr_\phi\rho^{\rm tot}_I(t)$ 
by adopting the \emph{projection operator method} (see, e.g., \cite{kubo}). 
We first introduce a linear operator 
$\PP\! : {\rm End}\,\cH^{\rm tot} \to {\rm End}\,\cH^{\rm tot}$  
which acts linearly on elements belonging to ${\rm End}\,\cH^{\rm tot}$ 
(the set of linear operators acting on $\cH^{\rm tot}$) 
and has the form 
\begin{align}
 \PP : \, O \, \mapsto \, \PP O \equiv (\Tr_{\phi}O) \otimes X^\phi
  \quad \bigl( O \in {\rm End}\,\cH^{\rm tot}\bigr)\,.
\end{align}
Here, $X^\phi \in {\rm End}\,\cH^{\phi}$ can be any operator acting on $\cH^\phi$ 
as long as it satisfies
\begin{align}
 \Tr_\phi X^\phi =1
\label{xphi}
\end{align}
and does not depend on time.
From \eqref{xphi} one can easily see that $\PP$ is a projection operator, 
$\PP^2=\PP$. 
For $O=\rho^{\rm tot}_I(t)$, 
we obtain 
\begin{align}
 \PP \rho^{\rm tot}_I(t) 
  = (\Tr_\phi \rho^{\rm tot}_I(t)) \otimes X^\phi
  =\rho_I(t) \otimes X^\phi \,.
\end{align}
We further introduce $\QQ\equiv1-\PP$\,, 
which is also a projection operator, $\QQ^2=\QQ$\,, 
and satisfies $\PP\QQ = 0 = \QQ\PP$.

Using the fact that operator $X^\phi$ can be chosen arbitrarily 
without changing the time evolution of $\rho_I(t)$ 
(as far as it satisfies the aforementioned conditions), 
we here set $X^\phi \equiv \rho^\phi(t_1) =\rho^\phi_I(t_1)$\,.
This certainly satisfies the condition \eqref{xphi},
and due to the initial condition \eqref{init_rho}
the following equations hold:
\begin{align}
 \PP\rho^{\rm tot}_I(t_1)=\rho^{\rm tot}_I(t_1)\,,\quad
 \QQ\rho^{\rm tot}_I(t_1)=0 \,.
 \label{pq_simple}
\end{align}
Then, if the one-point function of scalar field vanishes 
(as we assume hereafter), 
\begin{align}
\Tr_\phi\bigl( \phi_I\bigl(x(\tau)\bigr) \rho_I^\phi(t_1) \bigr) =0\,,
\label{1pt0}
\end{align} 
we obtain the equation 
\begin{align} 
 \frac{\rmd \rho_I(t)}{\rmd t} \otimes \rho^\phi_I(t_1) &=
 -\PP\ad_{V_I(t)}\int^{t}_{t_1}\rmd t'\,
 {\rm T} e^{-\ii\int^{t}_{t'}\rmd t^{\prime\prime}\QQ\ad_{V_I(t^{\prime\prime})}}
 \ad_{V_I(t')}\,\bigl( \rho_I(t') \otimes \rho^\phi_I(t_1) \bigr)\,.
\label{master_full}
\end{align}
We give a proof of Eq.~\eqref{master_full} in appendix \ref{app:master}.
This is the master equation 
which with the initial condition \eqref{init_rho} 
and the assumption \eqref{1pt0}
completely determines the time evolution of the reduced density matrix $\rho_I(t)$\,.

\subsection{Approximation of the master equation}
\label{subsec:master_approx}

We expand the right-hand side of \eqref{master_full} 
to the second order in perturbation to obtain
\begin{align}
 \frac{\rmd \rho_I(t)}{\rmd t} 
 =-\Tr_\phi \Big(\ad_{V_I(t)}\,\int^{t}_{t_1}\rmd t'
 \ad_{V_I(t')}\,\bigl( \rho_I(t') \otimes \rho^\phi_I(t_1) \bigr) \Bigr) 
 +\mathcal{O}(\lambda^3)\,.
\end{align}
This can be further rewritten in terms of the proper time $\tau$  
to the following form 
[denoting $\rho\bigl(t(\tau)\bigr)$ by $\rho(\tau)$ 
and derivatives with respect to $\tau$ by dots]:
\begin{align} 
 &\dot{\rho}_I(\tau)
\nn\\ 
  &= {}-\lambda^2 \int^{\tau}_{\tau_1}\!\!\rmd \tau'\, \Tr_\phi 
  \bigl[\mu_I(\tau) \otimes \phi_I\bigl(x(\tau)\bigr),\,
  \bigl[\mu_I(\tau')\otimes \phi_I\bigl(x(\tau')\bigr),\,  
  \rho_I(\tau') \otimes \rho^\phi_I(\tau_1) \bigr] \bigr] 
  +\mathcal{O}(\lambda^3)
\nn\\
 &= \lambda^2\int^{\tau}_{\tau_1}\rmd \tau' \Bigl(
 \bigl[\mu_I(\tau),\,\rho_I(\tau')\,\mu_I(\tau')\bigr]\,G_X^+\bigl(x(\tau'),x(\tau)\bigr)
\nn\\ 
 &\qquad -\bigl[\mu_I(\tau),\,\mu_I(\tau')\,\rho_I(\tau')\bigr]\,
 G_X^+\bigl(x(\tau),x(\tau')\bigr)
\Bigr) + \mathcal{O}(\lambda^3)\,,
\label{dense-int}
\end{align}
where
\begin{align}
 G_X^+(x,x')\equiv \Tr_\phi 
  \bigl( \phi_I(x)\,\phi_I(x')\,\rho^\phi_I(\tau_1) \bigr)
  \label{Wightman_X}
\end{align}
is the Wightman function of free scalar field 
with respect to the density matrix $X^\phi= \rho^\phi_I(\tau_1)$\,.

Since the reduced density matrix in the interaction picture, $\rho_I(\tau)$, 
is related to that in the Schr\"odinger picture, $\rho(\tau)$, 
as
\begin{align}
 \rho_I(\tau)= e^{\ii H^{\rm d}\cdot\,(\tau-\tau_1)}\,
 \rho(\tau)\,e^{{}-\ii H^{\rm d}\cdot\,(\tau-\tau_1)}\,,
\end{align}
we can rewrite \eqref{dense-int} to the following form in the Schr\"odinger picture: 
\begin{align} 
 \dot{\rho}(\tau) +\ii\,[H^{\rm d},\rho(\tau)]
 &=\lambda^2\int^{\tau}_{\tau_1}\rmd \tau' \bigl[
 \mu\, e^{{}-\ii H^{\rm d}\cdot\,(\tau-\tau')}\,\rho(\tau')\,\mu\,
 e^{\ii H^{\rm d}\cdot\,(\tau-\tau')}\,
 G_X^+\bigl(x(\tau'),x(\tau)\bigr)
\nn\\
 &\qquad\qquad
 + e^{{}-\ii H^{\rm d}\cdot\,(\tau-\tau')}\,\mu\,\rho(\tau')\,\mu\, 
 e^{\ii H^{\rm d}\cdot\,(\tau-\tau')}\,
 G_X^+\bigl(x(\tau),x(\tau')\bigr)
\nn\\
 &\qquad\qquad
 - e^{{}-\ii H^{\rm d}\cdot\,(\tau-\tau')}\,\rho(\tau')\,\mu\,
 e^{\ii H^{\rm d}\cdot\,(\tau-\tau')}\,\mu\,
 G_X^+\bigl(x(\tau'),x(\tau)\bigr)
\nn\\
 &\qquad\qquad
 -\mu\, e^{{}-\ii H^{\rm d}\cdot\,(\tau-\tau')}\,\mu\,\rho(\tau')\,
 e^{\ii H^{\rm d}\cdot\,(\tau-\tau')}\,
 G_X^+\bigl(x(\tau),x(\tau')\bigr)
\bigr]
\nn\\
&\quad + \mathcal{O}(\lambda^3)\,.
\label{dense'}
\end{align}

The integro-differential equation \eqref{dense'} can be further simplified 
as follows. 
Since the Wightman function $G_X^+\bigl(x(\tau),x(\tau')\bigr)$ in the integral 
is singular at $\tau'=\tau$ 
and decreases exponentially for large separations of $\tau$ and $\tau'$\,,%
\footnote{
We see in appendix \ref{app:Wightman-function} 
that for a scalar field in de Sitter space 
the Wightman function certainly exhibits this property 
if the mass is large enough. 
In general, there can be a case 
where the Wightman function has a long tail 
and one needs to take account of memory effects carefully.  
We do not deal with such cases in the present paper. 
}
the main contributions to the integral should come 
only from the region $\tau'\sim\tau$\,.  
This implies that the memory effect in the equation is highly suppressed, 
and thus we may replace $\rho_{kl}(\tau')$ in the integral 
by its boundary value $\rho_{kl}(\tau)$ to a good accuracy, 
assuming that $\rho(\tau')$ slowly changes. 
Equation \eqref{dense'} can thus be rewritten (in terms of matrix elements) as 
\begin{align}
 &\dot{\rho}_{mn}(\tau) +\ii(E_m-E_n)\rho_{mn}(\tau)
\nn\\
 &\simeq \lambda^2\sum_{k,l} \int^{\tau}_{\tau_1}\rmd\tau' \bigl[\,
 e^{-\ii(E_n-E_k)(\tau'-\tau)}\mu_{mk}\,\mu_{ln}\,  
 \rho_{kl}(\tau)\,G_X^+\bigl(x(\tau'),x(\tau)\bigr)
\nn\\
 &\qquad\qquad\quad\quad
 +e^{-\ii(E_m-E_l)(\tau-\tau')}\mu_{mk}\,\mu_{ln}\,
 \rho_{kl}(\tau)\,G_X^+\bigl(x(\tau),x(\tau')\bigr)
\nn\\
 &\qquad\qquad\quad\quad
 -e^{-\ii(E_l-E_m)(\tau'-\tau)}\mu_{kl}\,\mu_{ln}\,
 \rho_{mk}(\tau)\,G_X^+\bigl(x(\tau'),x(\tau)\bigr)
\nn\\
 &\qquad\qquad\quad\quad
 -e^{-\ii(E_k-E_n)(\tau-\tau')}\mu_{mk}\,\mu_{kl}\,
 \rho_{ln}(\tau)\,G_X^+\bigl(x(\tau),x(\tau')\bigr)
 \,\bigr]\,.
\label{dense}
\end{align}
If the off-diagonal elements of $\rho(\tau)$ 
can be further neglected,%
\footnote{
We will see in section \ref{sec:thermal} 
that we can set the off-diagonal elements of $\rho(\tau)$ 
can be set to zero without losing generality  
if the detector is a two-level system 
and $\mu$ has an off-diagonal form.}
then \eqref{dense} becomes
\begin{align} 
 \dot{\rho}_{mm}(\tau)
&= \sum_{k\neq m} \bigl[w^X_{mk}(\tau,\tau_1)\rho_{kk}(\tau)
 -w^X_{km}(\tau,\tau_1)\rho_{mm}(\tau)\bigr]\,.
\label{master}
\end{align}
Here, we have introduced the transition rate matrix 
\begin{align}
 w^X_{mk}(\tau,\tau_1) \equiv \lambda^2\,\abs{\mu_{mk}}^2 \,
 \dot{\cF}_X(E_m-E_k;\tau,\tau_1)\,,
\label{wmk}
\end{align}
where
\begin{align} 
 \dot{\cF}_X(\Delta E;\tau,\tau_1)&\equiv\int^{\tau}_{\tau_1}\rmd\tau' \bigl[\,
 e^{-\ii\Delta E(\tau'-\tau)} \,G_X^+\bigl(x(\tau'),x(\tau)\bigr)
 + e^{-\ii\Delta E(\tau-\tau')} \,G_X^+\bigl(x(\tau),x(\tau')\bigr)
 \bigr]\,. 
\label{fdot}
\end{align}
Equation \eqref{master} now has the standard form of the master equation.

If, in particular, 
$\displaystyle \dot\cF^{\rm eq}(\Delta E)\equiv 
\lim_{\tau-\tau_1\to\infty} \dot{\cF}_X(\Delta E;\tau,\tau_1)$ 
satisfies the relation%
\footnote{
The relation \eqref{thermaltransit} indeed holds 
with $\beta=2\pi\ell$ 
when $\rho^\phi(t_1)$ corresponds to the Bunch-Davies vacuum 
in de Sitter space. 
See Eq.~\eqref{equib_relation}.}
\begin{align}
 \frac{\dot{\cF}^{\rm eq}(\Delta E)}{\dot{\cF}^{\rm eq}(-\Delta E)}
  =e^{-\beta \Delta E}\,,
\label{thermaltransit}
\end{align}
the transition rate matrix satisfies the relation 
\begin{align}
 \frac{w_{mk}(\tau;\tau_1)}{w_{km}(\tau;\tau_1)}
  \,\overset{\tau-\tau_1\to\infty}{\longrightarrow} \,e^{{}-\beta\, (E_m-E_k)}\,.
\end{align}
Then, the distribution of the detector in equilibrium, $\rho^{\rm eq}$\,, 
may be determined by the detailed balance condition 
$\displaystyle\lim_{\tau-\tau_1\to\infty}w_{m k}(\tau;\tau_1)\,\rho^{\rm eq}_{k k}
=\displaystyle\lim_{\tau-\tau_1\to\infty}w_{k m}(\tau;\tau_1)\,\rho^{\rm eq}_{m m}$\,, 
and we obtain  
\begin{align}
 \rho^{\rm eq}_{mm}=\frac{e^{-\beta E_m}}{Z}\quad\bigl(Z=\sum_n e^{-\beta E_n}\bigr)\,,
\label{eq:Gibbs}
\end{align}
which is nothing but the Gibbs distribution at temperature $1/\beta$.

We close this section 
by making a comment on the relationship between our formalism and the literature. 
One can easily show that the transition rate matrix $w^X_{mk}$\,,
\eqref{wmk}, is the $\tau$-derivative of 
\begin{align}
 \lambda^2\, \abs{\mu_{mk}}^2\, \cF_X(E_m-E_k;\tau,\tau_1)\,,
\label{transition_prob}
\end{align}
where 
\begin{align}
 \cF_X(\Delta E;\tau,\tau_1)=
  \int^{\tau}_{\tau_1}\rmd \tau'\int^{\tau}_{\tau_1}\rmd \tau^{\prime\prime} 
  e^{-\ii\Delta E(\tau'-\tau^{\prime\prime})}\,
  G_X^+\bigl(x(\tau'),x(\tau^{\prime\prime})\bigr)\,.
\label{response}
\end{align}
In the literature (e.g.\ \cite{Birrell:1982ix}), 
one often considers a process 
from an initial state $\ket{k}\otimes\ket{\alpha}$ at time $t_1$ 
(usually taken to be the infinite past)
to a final state $\ket{m} \otimes \ket{\beta}$ at time $t$\,, 
and sums over the final states $\ket{\beta}$ of the scalar field. 
The transition probability has the same form 
as \eqref{transition_prob} 
if we set $X^\phi=\rho^\phi(t_1)=\rho_I^\phi(t_1)=\ket{\alpha}\bra{\alpha}$ 
for which the Wightman function becomes 
$G_X^+(x,x')=\bra{\alpha}\,\phi_I(x)\,\phi_I(x')\,\ket{\alpha}$\,. 
We thus again see that \eqref{wmk} represents the transition probability 
per unit proper time of the detector. 
$\cF_X$ in \eqref{response} is often called 
the response function (see, e.g., \cite{Birrell:1982ix}). 
As we have seen, $\cF_X$ or its derivative $\dot{\cF}_X$ 
does not depend on details of the detector 
and can be thoroughly determined by the Wightman function of free scalar field.

\section{Unruh-DeWitt detector in de Sitter space}
\label{sec:dS}

In this section we consider an Unruh-DeWitt detector 
in $d$-dimensional de Sitter space, 
which is weakly interacting with a massive scalar field $\phi(x)$ 
of mass $m$\,.  
We exclusively consider the Poincar\'e patch 
denoting the time variable by $\eta$\,,
\begin{align}
 \rmd s^2 = \ell^2\,\frac{{}-\rmd \eta^2 + \rmd \bx^2}{\eta^2}
  \quad \bigl(-\infty < \eta < 0\bigr)\,,
\end{align}
and set the classical trajectory of the detector to be the geodesic
\begin{align}
 x^{\mu}(\tau) = \bigl({}-\ell\, e^{-\tau/\ell},\,{\mathbf0}\bigr)\,.
\label{eq:geodesic}
\end{align}
We denote by $\eta_1$ 
the time when the detector 
starts the interaction with the field $\phi(x)=\phi(\eta,\bx)$, 
which has the following form in the interaction picture 
[see \eqref{vint}]:
\begin{align}
 V_I(\eta)=\lambda\,\mu_I(\tau)\, \frac{\rmd \tau(\eta)}{\rmd \eta}
  \otimes\phi_I\bigl(x(\tau)\bigr)\,
  \theta(\eta-\eta_1)\,.
\end{align}

We will show that the density distribution of the detector 
exhibits a relaxation to the Gibbs distribution with $\beta=2\pi\ell$ 
when the initial condition $X^\phi=\rho^\phi_I(\eta_1)$ satisfies the condition \eqref{1pt0} 
and the corresponding Wightman function $G_X^+(x,x')$ 
has the same short distance behavior ($\abs{\bx-\bx'}\to 0$ with $\eta=\eta'$) 
as that of the Bunch-Davies vacuum, $G_{\rm BD}^+(x,x')$.
In the following, we set the curvature radius $\ell=1$.

Let the initial density matrix $X^\phi=\rho_I^\phi(\eta_1)$ of scalar field have the form
\begin{align}
 X^\phi=\ket{\rm BD}\bra{\rm BD} + \Delta X^\phi\,,
\end{align}
for which the Wightman function takes the form 
\begin{align}
 G_X^+(x,x')=\Tr_\phi\bigl(\phi_I(x) \phi_I(x') X^\phi\bigr)
 =G_{\rm BD}^+(x,x') + \Delta G^+(x,x')\,.
\label{Wightman-decomposition}
\end{align}
We assume that the deformed Wightman function is invariant 
under spatial translations and rotations  
for fixed $\eta$ and $\eta'$\,, 
and write its Fourier transform as ($k\equiv |\bk|$)
\begin{align}
 G_X^+(\eta,\bx,\eta',\bx')
 &\equiv \int\!\frac{\rmd^{d-1}\bk}{(2\pi)^{d-1}}\,e^{\ii\,\bk\cdot(\bx-\bx')}\,
 G_{X,k}^+(\eta,\eta')
\nn\\
 &\equiv \int\!\frac{\rmd^{d-1}\bk}{(2\pi)^{d-1}}\,e^{\ii\,\bk\cdot(\bx-\bx')}\,
 \bigl[G^+_{{\rm BD},k}(\eta,\eta') + \Delta G_{k}^+(\eta,\eta')\bigr]\,.
\end{align}
This takes the following form for the geodesic \eqref{eq:geodesic} 
[with $\eta=-e^{-\tau}$ and $\eta'=-e^{-\tau'}$]:
\begin{align}
 G_X^+\bigl(x(\tau),x(\tau')\bigr)
 &=\int\!\frac{\rmd^{d-1}\bk}{(2\pi)^{d-1}}\,G_{X,k}^+(\eta,\eta')
\nn\\
 &=\frac{2}{(4\pi)^{\frac{d-1}{2}}\,\Gamma\bigl(\frac{d-1}{2}\bigr)}\,\int_0^\infty\!\rmd k\,k^{d-2}\,
 G_{X,k}^+(\eta,\eta')
\nn\\
 &\equiv G_{\rm BD}^+\bigl(x(\tau),x(\tau')\bigr)
 +\Delta G^+\bigl(x(\tau),x(\tau')\bigr)\,.
\end{align}

Note that both $G_{X,k}^+(\eta,\eta')$ and $G^+_{{\rm BD},k}(\eta,\eta')$ 
[and thus $\Delta G_{k}^+(\eta,\eta') $ also] 
satisfy the homogeneous Klein-Gordon equation, 
$\eta^2\,\ddot{f}(\eta)+(d-2)\,\eta\,\dot{f}(\eta)+(k^2\,\eta^2+m^2)\,f(\eta)=0$\,, 
with respect to each of the arguments $\eta$ and $\eta'$\,. 
The solutions to this equation are given by linear combinations of 
$ (-\eta)^{(d-1)/2}\,H_\nu^{(a)}(-k\eta)$ $(a=1,2)$\,,
where $H_\nu^{(1,2)}(z)$ are the Hankel functions 
and 
\begin{align}
  \nu &\equiv \left\{ 
  \begin{array}{ll}
 \displaystyle
   \sqrt{\Bigl(\frac{d-1}{2}\Bigr)^2- m^2} 
     &\displaystyle \Bigl(0< m < \frac{d-1}{2}\Bigr)
\\[3mm]
 \displaystyle
   \ii\,\sqrt{m^2 - \Bigl(\frac{d-1}{2}\Bigr)^2} \quad 
     &\displaystyle \Bigl(m \geq \frac{d-1}{2}\Bigr) 
  \end{array}
  \right..
\end{align}
Thus the Wightman function is generically given by a linear combination of 
\begin{align}
 [(-\eta)(-\eta')]^{\frac{d-1}{2}}\,
 H^{(a)}_{\nu}(-k\,\eta) \,H^{(b)}_{\nu}(-k\,\eta')
 \quad (a,b=1,2)\,. 
\end{align}
For example, $G^+_{{\rm BD},k}(\eta,\eta')$ is given by 
(see \cite{Bunch:1978yq,Birrell:1982ix,Fukuma:2013mx})
\begin{align} 
\label{BD_Wightman}
 G^+_{{\rm BD},k}(\eta,\eta')
 &= 
 \frac{\pi}{4}\,\bigl[(-\eta)\,(-\eta')\bigr]^{\frac{d-1}{2}} \,
 H^{(1)}_\nu(-k\,\eta)\,H^{(2)}_\nu(-k\,\eta')  \,.
\end{align}
Due to the condition that $G_{X}^+(x,x')$ and $G_{{\rm BD}}^+(x,x')$ 
have the same short distance behavior, 
$\Delta G^+_k(\eta,\eta')$ takes the following form 
($c_0$ is a constant with the dimension of time):
\begin{align}
 \Delta G_{k}^+(\eta,\eta') 
 = [(-\eta)(-\eta')]^{\frac{d-1}{2}}\,
  \sum_{a,b=1}^2 f_{ab}(c_0\,k)\, H^{(a)}_{\nu}(-k\,\eta) \,H^{(b)}_{\nu}(-k\,\eta')
\label{eq:DeltaG+}
\end{align}
with%
\footnote{
$\alpha_{ab}$ will need to be integers 
when imposing analyticity on the Wightman functions. }
\begin{align}
 f_{ab}(z) = {\rm const.} \, z^{-\alpha_{ab}}\,\bigl[1+\mathcal{O}(z^{-1})\bigr] \qquad 
 \bigl(\alpha_{ab}>0 \bigr) \,.
\label{f_ab-alpha}
\end{align}
Note that we only require that the leading singularities 
be the same for $G_{X}^+(x,x')$ and $G_{\rm BD}^+(x,x')$\,, 
and the functions $f_{ab}(z)$ control the subleading singularities. 

The derivative of the response function, $\dot{\cF}_X$ 
[defined in Eq.~\eqref{fdot}], can now be written as 
\begin{align}
 \dot{\cF}_X(\Delta E;\tau,\tau_1) = 
 \dot{\cF}_{\rm BD}(\Delta E;\tau,\tau_1)+\Delta\dot{\cF}(\Delta E;\tau,\tau_1)\,,
\end{align}
where
\begin{align}
 \dot{\cF}_{\rm BD}(\Delta E;\tau,\tau_1)
 &\equiv \int_{-(\tau-\tau_1)}^0\rmd s\,
  e^{-\ii\Delta E\,s}\,G^{+}_{\rm BD}\bigl(x(\tau+s),x(\tau)\bigr)
\nn\\ 
 &\quad +\int_0^{(\tau-\tau_1)}\rmd s\,
  e^{-\ii\Delta E\,s}\,G^{+}_{\rm BD}\bigl(x(\tau),x(\tau-s)\bigr)\,,
  \label{fdot_BD_int}
\\
\Delta\dot{\cF}(\Delta E;\tau,\tau_1)
 &\equiv \int_{-(\tau-\tau_1)}^0\rmd s\,
  e^{-\ii\Delta E\,s}\,\Delta G^{+}\bigl(x(\tau+s),x(\tau)\bigr)
\nn\\ 
 &\quad +\int_0^{(\tau-\tau_1)}\rmd s\,
  e^{-\ii\Delta E\,s}\,\Delta G^{+}\bigl(x(\tau),x(\tau-s)\bigr)\,.
  \label{delta_fdot_int}
\end{align}
The integrals can be evaluated analytically 
as shown in appendix \ref{app:integration}, 
and we find that they take the following asymptotic forms 
in the limit $\tau \to \infty$ [see Eqs.~\eqref{eq:BD-asympt} and \eqref{eq:DF-asympt}]:
\begin{align}
 \dot{\cF}_{\rm BD}(\Delta E;\tau,\tau_1)
 &\sim \dot{\cF}^{\eq} (\Delta E)
 +{\rm const.} \, e^{-(\frac{d-1}{2}\pm\nu\pm\ii\Delta E)\,(\tau-\tau_1)}\,,
\label{BD_fdot_asympt}
\\
 \Delta\dot{\cF}(\Delta E;\tau,\tau_1) 
 &\sim {\rm const.}\,e^{-\alpha\,\tau} 
  + {\rm const.}\,e^{-(\frac{d-1}{2}\pm\nu\pm\ii\Delta E)\,(\tau-\tau_1)}
\label{delta_fdot_asympt}
\end{align}
with $\alpha \equiv \underset{a,b}{\rm min}(\alpha_{ab})$\,.

The $\tau$-independent term $\dot{\cF}^{\eq} (\Delta E)$ 
is given by [see Eq.~\eqref{F_eq}]
\begin{align}  
 \dot{\cF}^{\rm eq}(\Delta E)
 &= \frac{e^{-\pi\ell\,\Delta E}\, 
 \Gamma\Bigl(\frac{\frac{d-1}{2}+\nu+\ii \ell\Delta E}{2}\Bigr)\,
 \Gamma\Bigl(\frac{\frac{d-1}{2}-\nu+\ii \ell\Delta E}{2}\Bigr)\,
 \Gamma\Bigl(\frac{\frac{d-1}{2}+\nu-\ii \ell\Delta E}{2}\Bigr)\,
 \Gamma\Bigl(\frac{\frac{d-1}{2}-\nu-\ii \ell\Delta E}{2}\Bigr)}
         {8\,\pi^{\frac{d+1}{2}}\,\Gamma\bigl(\frac{d-1}{2}\bigr)}\,,
 \label{eq:Fdot-eq}
\end{align}
which agrees with the known result 
obtained in \cite{Garbrecht:2004du,Higuchi:1986ww} 
(we have restored the curvature radius $\ell$). 
One easily finds that this satisfies the relation 
\begin{align}
 \frac{\dot{\cF}^{\rm eq}(\Delta E)}{\dot{\cF}^{\rm eq}(-\Delta E)}
  =e^{-2\pi\ell \Delta E}\,.
\label{equib_relation}
\end{align}
Thus, from the argument following \eqref{thermaltransit},
we confirm that, as $\tau$ becomes large,  
the density distribution $\rho_{mm}(\tau)$ approaches the Gibbs distribution 
at temperature $1/2\pi\ell$\,,
\begin{align}
 \rho^{\eq}_{mm}
 = \frac{e^{-2\pi\ell E_m}}{Z} \quad
 \bigl(Z=\sum_n e^{-2\pi\ell E_n}\bigr)\,. 
\end{align}

The $\tau$-dependent terms, $e^{-\alpha\,\tau/\ell}$ 
and $e^{-[(\frac{d-1}{2}\pm\nu)/\ell\pm\ii\Delta E]\,\tau}$\,,
in the asymptotic forms \eqref{BD_fdot_asympt} and \eqref{delta_fdot_asympt}
represent relaxation modes of the detector 
with relaxation times $\ell/\alpha$ and $\abs{(d-1)/2\pm {\rm Re}\,\nu}^{-1}\ell$, 
respectively. 
Note that the modes with the latter relaxation time 
exist even when the initial state is the Bunch-Davies vacuum, 
in which we are not interested here. 
On the other hand, the mode with relaxation time $\ell/\alpha$ arises 
only when the initial state of scalar field 
is deviated from the Bunch-Davies vacuum. 
We thus identify with $\ell/\alpha$ 
the relaxation time 
for the surrounding medium [see Fig.~\ref{fig:medium} (b)] 
which relaxes from a nonequilibrium state 
to the equilibrium of temperature $1/2\pi\ell$\,.

We close this section with writing down the full expression of \eqref{BD_fdot_asympt} 
for comparison with the known results in the literature [see Eq.~\eqref{eq:BD-full}]:
\begin{align} 
 &\dot{\cF}_{\rm BD}(\Delta E,\tau,\tau_1)
\nn\\
 &= \dot\cF^{\rm eq}(\Delta E)
\nn\\ 
  &\ +\frac{e^{\ii\pi\,\frac{d-1}{2}}\,e^{\ii\pi\,\nu}\,
                   e^{-(\frac{d-1}{2}+\nu-\ii\Delta E)\,(\tau-\tau_1)}}
                  {8\,\pi^{\frac{d-1}{2}}\,\Gamma\bigl(\frac{d-1}{2}\bigr)\,\sin(\pi\,\nu)} \,
  \, _3\widehat{F}_2\Biggl(
  \afrac{\frac{d-1}{2},\,
                      \frac{\frac{d-1}{2}+\nu-\ii\Delta E}{2} ,\, 
                      \frac{d-1}{2}+\nu}
                     {1+\nu,\,
                      \frac{\frac{d+3}{2}+\nu-\ii\Delta E}{2}};\, 
                      e^{-2 (\tau-\tau_1)} \Biggr) 
\nn\\ 
 & \ +\frac{e^{\ii\pi\,\frac{d-1}{2}}\,e^{-\ii\pi\,\nu}\,
                 e^{-(\frac{d-1}{2}-\nu-\ii\Delta E)\,(\tau-\tau_1)}}
                {8\,\pi^{\frac{d-1}{2}}\,\Gamma\bigl(\frac{d-1}{2}\bigr)\,\sin(-\pi\,\nu)} \,
  \, _3\widehat{F}_2\Biggl(
  \afrac{\frac{d-1}{2},\,
                      \frac{\frac{d-1}{2}-\nu-\ii\Delta E}{2},\, 
                      \frac{d-1}{2}-\nu}
                     {1-\nu,\,
                      \frac{\frac{d+3}{2}-\nu-\ii\Delta E}{2}};\, 
                     e^{-2 (\tau-\tau_1)}\Biggr)
\nn\\ 
 & \ +\frac{e^{-\ii\pi\,\frac{d-1}{2}}\,e^{-\ii\pi\,\nu}\,
                   e^{-(\frac{d-1}{2}+\nu+\ii\Delta E)\,(\tau-\tau_1)}}
                  {8\,\pi^{\frac{d-1}{2}}\,\Gamma\bigl(\frac{d-1}{2}\bigr)\,\sin(\pi\,\nu)} \,
  \, _3\widehat{F}_2\Biggl(
  \afrac{\frac{d-1}{2},\,
                      \frac{\frac{d-1}{2}+\nu+\ii\Delta E}{2} ,\, 
                      \frac{d-1}{2}+\nu}
                     {1+\nu,\,
                      \frac{\frac{d+3}{2}+\nu+\ii\Delta E}{2}};\, 
                     e^{-2 (\tau-\tau_1)} \Biggr)  
\nn\\ 
 & \ +\frac{e^{-\ii\pi\,\frac{d-1}{2}}\,e^{\ii\pi\,\nu}\,
                 e^{-(\frac{d-1}{2}-\nu+\ii\Delta E)\,(\tau-\tau_1)}}
                {8\,\pi^{\frac{d-1}{2}}\,\Gamma\bigl(\frac{d-1}{2}\bigr)\,\sin(-\pi\,\nu)} \,
  \, _3\widehat{F}_2\Biggl(
  \afrac{\frac{d-1}{2},\,
                      \frac{\frac{d-1}{2}-\nu+\ii\Delta E}{2} ,\, 
                      \frac{d-1}{2} -\nu}
                     {1-\nu,\,
                      \frac{\frac{d+3}{2}-\nu+\ii\Delta E}{2}};\, 
                     e^{-2 (\tau-\tau_1)}\Biggr) \Biggr]\,.
\label{eq:Fdot-n=0}
\end{align} 
If we set $d=4$ and $\nu=1/2$, 
Eq.~\eqref{eq:Fdot-n=0} correctly reproduces 
the result obtained in section~IV-D of \cite{Garbrecht:2004du}, 
where the case $\tau_0=-\infty$ is considered 
so that the higher-order corrections 
$\Delta\dot{\cF}(\Delta E,\tau,\tau_1)$ 
are all zero.

\section{The case of the instantaneous ground state at a finite past}
\label{sec:fintie_past}

In this section, 
we consider the case where the initial state $X^\phi=\rho^\phi_I(\eta_1)$ 
is the instantaneous  ground state at an earlier time $\eta_0$\,.
In this case, $\Delta\dot{\cF}(\Delta E;\tau,\tau_1)$ 
can be calculated explicitly, 
and we will see that the relaxation time takes a universal value $\ell/2$ 
(i.e., $\alpha=2$), 
irrespectively of the values of $\eta_0$ and $\eta_1$ 
or the form of interaction between the detector and the scalar.

\subsection{The instantaneous ground states}
\label{subsec:dS_instantaneous}

\begin{figure}[htbp]
\begin{center}
\includegraphics[width=5.5cm]{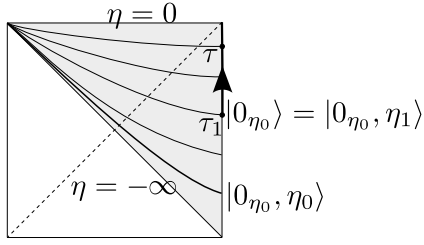}
\begin{quote}
\caption{The trajectory of an Unruh-DeWitt detector in the Penrose diagram 
of de Sitter space. 
The shaded region corresponds to the Poincar\'e patch, 
and the dashed line represents the future event horizon for the detector. 
The scalar field is in the ground state at time $\eta_0$\,. 
\label{fig:Poincare-detector}
}
\end{quote}
\end{center}
\vspace{-3ex}
\end{figure}

Let $\ket{0_{\eta_0}, \eta_0}$ denote the Schr\"odinger picture state 
which is the instantaneous ground state at an early time $\eta_0$ 
(see Fig.~\ref{fig:Poincare-detector}). 
We then introduce the state $\ket{0_{\eta_0}}\equiv\ket{0_{\eta_0},\eta_1}$ 
as an interaction picture state 
(or as a Heisenberg picture state of free field theory) 
which is obtained by applying the free-field time evolution operator 
from $\eta_0$ to $\eta_1$\,:  
$\ket{0_{\eta_0}}=
{\rm T}\exp\bigl(-\ii\int_{\eta_0}^{\eta_1}\!\!\rmd \eta'\,H^\phi(\eta')\bigr)\,
\ket{0_{\eta_0},\eta_0}$\,. 
In this section, we set $X^\phi=\rho^\phi_I(\eta_1)=\ket{0_{\eta_0}} \bra{0_{\eta_0}}$\,. 
See \cite{Fukuma:2013mx} 
for a detailed discussion on defining such time-dependent ground states 
for free scalar field in curved spacetimes.

Note that the one-point function vanishes, 
$\Tr_\phi \bigl(\phi_I(x)X^\phi\bigr)
=\bra{0_{\eta_0}}\,\phi_I(x)\,\ket{0_{\eta_0}}=0$\,. 
Then, with the approximations made in subsection \ref{subsec:master_approx}, 
we have the master equation of the form [see \eqref{master}--\eqref{fdot}]
\begin{align}
 \dot{\rho}_{mm}(\tau)
 =\sum_{k\neq m} \bigl[w_{mk}(\tau,\tau_1;\eta_0)\,\rho_{kk}(\tau)
 -w_{km}(\tau,\tau_1;\eta_0)\,\rho_{mm}(\tau)\bigr] 
\label{master_P}
\end{align}
with the transition rate matrix 
\begin{align}
 w_{mk}(\tau,\tau_1;\eta_0) \equiv \lambda^2\,\abs{\mu_{mk}}^2 \,
 \dot{\cF}(E_m-E_k;\tau,\tau_1;\eta_0)\,.
\label{wmk_P}
\end{align}
Here, the derivative of the response function is given by 
\begin{align}
 \dot{\cF}(\Delta E;\tau,\tau_1;\eta_0)
 &\equiv\int^{\tau}_{\tau_1}\rmd\tau' \bigl[\,
 e^{-\ii\Delta E(\tau'-\tau)} \,G^+\bigl(x(\tau'),x(\tau);\eta_0\bigr)
\nn\\
 &\qquad\qquad + e^{-\ii\Delta E(\tau-\tau')} \,G^+\bigl(x(\tau),x(\tau');\eta_0\bigr)
 \bigr] 
\label{fdot_P}
\end{align}
with the Wightman function 
\begin{align}
 G^+(x,x';\eta_0)\equiv \bra{0_{\eta_0}}\phi_I(x)\phi_I(x')\ket{0_{\eta_0}}\,.
\label{Wightman_P} 
\end{align}
In the following, we compute $G^+(x,x';\eta_0)$ for finite $\eta_0$ 
using a technique developed in \cite{Fukuma:2013mx}, 
and will find that $G^+(x,x';\eta_0)$ indeed has the form \eqref{Wightman-decomposition} 
and \eqref{eq:DeltaG+} 
with
\begin{align}
 f_{12}(z)=f_{21}(z)= 
 \frac{\pi}{4}\Bigl(\frac{d-2}{4}\Bigr)^2\,z^{-2}+{\cal O}(z^{-4})\,,
 \quad f_{11}(z)=f_{22}(z)={\cal O}(z^{-4})\,.
\label{eq:fab}
\end{align} 
Thus, $\alpha=2$ in this model, 
and the relaxation time of a medium surrounding the detector 
is given by $\ell/2$\,. 
Since an explicit functional form can be obtained for $\Delta G^+(x,x';\eta_0)$ in this model, 
we can further determine the coefficients of the damping term $e^{-\alpha\,\tau}$  
in Eq.~\eqref{delta_fdot_asympt}.

We comment that the instantaneous ground states $\ket{0_{\eta_0}}$ 
differ from the adiabatic vacuum state 
\cite{Parker:PhD,Parker:1974qw} (see also \cite{Birrell:1982ix,PT}) 
at second adiabatic order 
for finite $\eta_0$\,. 
This rapid departure may be expected from the fact that 
the condition of the validity of adiabatic approximation 
[$\omega_k \gg \dot{\omega_k}/\omega_k$ with $\omega_k$ given by \eqref{omega_k}] 
does not hold for some modes $\bk$ if $\eta_0$ is finite. 
Actually, while the adiabatic vacuum state has the same asymptotic expansion 
as the Bunch-Davies vacuum, 
the wave function of an instantaneous ground state 
agrees with that of the Bunch-Davies vacuum 
only at the leading asymptotic order [see \eqref{wave-function_asympto}]. 
This observation then means that we cannot use the adiabatic subtraction scheme 
to regularize the expectation value of the energy-momentum tensor, 
$\langle T_{\mu\nu}\rangle$\,, 
for instantaneous ground states. 
In fact, 
it is shown in \cite{Anderson:2000wx, Anderson:2005hi} 
that, in order for a state in $d$-dimensional de Sitter space 
to admit such adiabatic regularization of $\langle T_{\mu\nu}\rangle$\,, 
the state must be of $d$-th or higher adiabatic order; 
in other words, the wave function $\varphi_k(\eta)$ of the state must have the form 
\begin{align}
 \varphi_k(\eta)=(-\eta)^{\frac{d-1}{2}}\,\bigl[c_1(k)\,H^{(1)}_\nu(-k \eta)
 +c_2(k)\,H^{(2)}_\nu(-k \eta) \bigr]
\end{align}
with $c_1(k)={\rm const.}+o(k^{-d})$ and $c_2(k)=o(k^{-d})$\,,%
\footnote{
For such states, 
the exponents $\alpha_{ab}$ in \eqref{f_ab-alpha} are greater than $d$\,, 
and thus the relaxation time will take a value less than $\ell/d$\,. 
}
which is clearly different from the form given in \eqref{wave-function_asympto}. 
We expect that one can yet use the point-splitting regularization 
to remove divergences of $\langle T_{\mu\nu}\rangle$ for instantaneous ground states 
although one then needs to establish an algorithm 
to determine the finite part of $\langle T_{\mu\nu}\rangle$\,. 
This issue is beyond the scope of this paper and is left for the future work.

\subsection{Wightman function for the instantaneous ground state}
\label{subsec:dS_Wightman}

In the Poincar\'e patch, there is a translational invariance in the spatial directions, 
and we assume that all spatial directions are compactified with radius $L/2\pi$.%
\footnote{ 
The introduction of $L$ is for dealing with the zero mode carefully. 
$L$ will be taken to infinity in the end. }
Then the wave vector $\bk$ takes discrete values 
$\bk =(2\pi/L)\,\bn$ with $\bn\in\bZZ^{d-1}$\,.
We denote $\bk>0$ (or $\bk<0$) 
if the first nonvanishing element of the vector 
$\bk=(k_1,\,k_2,\,\ldots\,,k_{d-1})$ is positive (or negative). 
Now, we consider the mode expansion
\begin{align}
 \phi(x) = \phi(\eta,\bx) = \sum_{\bk\geq0}\,\sum_{a} \phi_{\bk,a}(\eta)\, Y_{\bk,\,a}(\bx) \,,
\label{eq:mode-expansion}
\end{align}
where the mode functions $\bigl\{Y_{\bk,\,a}(\bx)\bigr\}$ are given by 
\begin{align}
 \bk=0:\quad&Y_{\bk=0,\,a=1}\equiv\frac{1}{\sqrt{V}}\quad
  \bigl(V\equiv L^{d-1}\bigr)\,, \\
 \bk>0:\quad&
  Y_{\bk,\, a=1}(\bx)=\sqrt{\dsfrac{2}{V}}\,\cos(\bk\cdot\bx)\,,\quad
  Y_{\bk,\, a=2}(\bx)=\sqrt{\dsfrac{2}{V}}\,\sin(\bk\cdot\bx)\,,
\end{align}
which form a complete set of (real-valued) 
eigenfunctions of the spatial Laplacian 
$\Delta_{d-1}=\sum_{i=1}^{d-1}\partial_i^2$\,.
Then, the Wightman function \eqref{Wightman_X} is given by
\begin{align}
 G_X^+(x,x') 
 &= \sum_{\bk\geq 0}\,G_{X,k}^+(\eta,\eta')\,
    \sum_a\,Y_{\bk,a}(\bx)\,Y_{\bk,a}(\bx')
\nn\\
 &= \frac{1}{(2\pi)^{\frac{d-1}{2}}\,\abs{\bx-\bx'}^{\frac{d-3}{2}}}\,
    \int_0^\infty \!\!\!\rmd k\,k^{\frac{d-1}{2}}\,
             J_{\frac{d-3}{2}}\bigl(k\,\abs{\bx-\bx'}\bigr)\, 
    G_{X,k}^+(\eta,\eta') 
\label{G+_general}
\end{align}
with
\begin{align}
 G_{X,k}^+(\eta,\eta') 
 = \bra{0_{\eta_0}} \phi_{\bk,a}(\eta)\,\phi_{\bk,a}(\eta') \ket{0_{\eta_0}}\,,
  \label{G+_mode}
\end{align}
where we have taken the limit $L\to\infty$ 
and integrated over angular variables 
in the second equality of Eq.~\eqref{G+_general}.

With the mode expansion \eqref{eq:mode-expansion}, 
the Hamiltonian (in the Heisenberg picture of free field theory)
is expressed as a sum of the Hamiltonians 
for independent harmonic oscillators:
\begin{align} 
 H(\eta) &= \sum_{\bk,a} 
   \omega_{k}(\eta)\,a_{\bk,a}^\dag(\eta)\,a_{\bk,a}(\eta) 
   + \mbox{const.}\,,
\\ 
 a_{\bk,a}(\eta) &\equiv \frac{(-\eta)^{-\frac{d-2}{2}}}{\sqrt{2}}\,\Bigl[\,
         \omega_{k}^{1/2}(\eta) \,\phi_{\bk,a}(\eta)
       +\ii\, \omega_{k}^{-1/2}(\eta)\,\dot{\phi}_{\bk,a}(\eta)\,\Bigr]\,,
\\ 
 \omega_k(\eta)&\equiv \sqrt{m^2 \,(-\eta)^{-2}+k^2} 
\label{omega_k}
\,.
\end{align}
The instantaneous ground state $\ket{0_{\eta_0}}$ 
is then defined by the condition 
that $a_{\bk,a}(\eta_0)\,\ket{0_{\eta_0}} = 0$ 
$(\forall\, \bk,\,a)$.

By expanding the operator $\phi_{\bk,a}(\eta)$ 
using the annihilation and creation operators $a_{\bk,a}(\eta_0)$ and $a^\dag_{\bk,a}(\eta_0)$ 
as
\begin{align}
 \phi_{\bk,a}(\eta) = \varphi_k(\eta;\eta_0)\,a_{\bk,a}(\eta_0)
                   +\varphi_k^*(\eta;\eta_0)\,a_{\bk,a}^\dag(\eta_0) \,,
\end{align}
the ground state wave function $\varphi_k(\eta;\eta_0)$ 
is given by (see \cite{Fukuma:2013mx})
\begin{align}
 \varphi_k(\eta;\eta_0)
  ={} -\frac{\pi}{2\sqrt{2\omega_k(\eta_0)}}\,
    \frac{(-\eta)^{\frac{d-1}{2}}}{(-\eta_0)^{-\frac{d-2}{2}}}\,
    \bigl[ v_0\,J_{\nu}(-k\,\eta) - u_0\,N_{\nu}(-k\,\eta) \bigr] \,,
\label{eq:wave-function}
\end{align}
where  
$J_\nu(x)$ and $N_\nu(x)$ are the Bessel and Neumann functions, respectively, 
and  
$u_0$ and $v_0$ are given by
\begin{align} 
 u_0
  &= -(-\eta_0)^{-\frac{d-1}{2}}\,
      \Bigl[\Bigl(\frac{d-1}{2}+\nu +\ii\omega_k(\eta_0)\,\eta_0\Bigr)\,
            J_{\nu}(-k\,\eta_0)
            +k\,\eta_0\,J_{1+\nu}(-k\,\eta_0)\Bigr] \,,
\\ 
 v_0
  &= -(-\eta_0)^{-\frac{d-1}{2}}\,
      \Bigl[\Bigl(\frac{d-1}{2}+\nu +\ii\omega_k(\eta_0)\,\eta_0\Bigr)\,
            N_{\nu}(-k\,\eta_0)
            +k\,\eta_0\,N_{1+\nu}(-k\,\eta_0)\Bigr] \,. 
\end{align}
If we expand the wave function $\varphi_k(\eta;\eta_0)$ 
around $\eta_0=-\infty$, 
we find that the Wightman function for each mode,
\begin{align}
 G_k^+(\eta,\eta';\eta_0) 
   = \bra{0_{\eta_0}}\,\phi_{\bk,\,a}(\eta)\,\phi_{\bk,\,a}(\eta')\,\ket{0_{\eta_0}}
   = \varphi_k(\eta;\eta_0)\,\varphi^*_k(\eta';\eta_0) \,, 
\label{mode_sum}
\end{align}
takes the following form (see appendix \ref{app:asympto_Wightman-function}): 
\begin{align} 
 &G^{+}_{k}(\eta,\eta';\eta_0) 
\nn\\ 
 &= \frac{\pi}{4}\,\bigl[(-\eta)\,(-\eta')\bigr]^{\frac{d-1}{2}} \,
 \Bigl\{H^{(1)}_\nu(-k\,\eta)\,H^{(2)}_\nu(-k\,\eta') 
\nn\\ 
 &\quad +\Bigl(\frac{d-2}{4}\Bigr)^2\,\frac{1}{(-k\,\eta_0)^2} \,
 \bigl[H^{(1)}_\nu(-k\,\eta)\,H^{(2)}_\nu(-k\,\eta')
               + H^{(2)}_\nu(-k\,\eta)\,H^{(1)}_\nu(-k\,\eta')\bigr]
\nn\\ 
 &\quad +
   e^{2\ii[k\,\eta_0+\frac{\pi(1+2\nu)}{4}]}\,\Bigl[\ii\frac{d-2}{4(-k\,\eta_0)}
      +\frac{\bigl(\nu^2-\frac{1}{4}\bigr)\,(d-3)
   -\bigl(\frac{d-2}{2}\bigr)^2-m^2}{4(-k\,\eta_0)^2}\Bigr]\,H^{(1)}_\nu(-k\,\eta)\,H^{(1)}_\nu(-k\,\eta')
\nn\\ 
 &\quad +
   e^{-2\ii[k\,\eta_0+\frac{\pi(1+2\nu)}{4}]}\,\Bigl[-\ii\frac{d-2}{4(-k\,\eta_0)}
      +\frac{\bigl(\nu^2-\frac{1}{4}\bigr)\,(d-3)
  -\bigl(\frac{d-2}{2}\bigr)^2-m^2}{4(-k\,\eta_0)^2}\Bigr] \,H^{(2)}_\nu(-k\,\eta)\,H^{(2)}_\nu(-k\,\eta')
\nn\\ 
 &\quad +\mathcal{O}((-k\,\eta_0)^{-3})\Bigr\} \,.
\label{Wightman_k_asympt}
\end{align}
Here, the first term, which can be written as 
$\displaystyle\lim_{\eta_0\to -\infty}G^{+}_{k}(\eta,\eta';\eta_0)$\,, 
is the Wightman function for the Bunch-Davies vacuum, 
$G^{+}_{{\rm BD},k}(\eta,\eta')$\,. 
The last two terms include rapidly oscillating factors 
$e^{\pm 2\ii k\,\eta_0}$ 
and can be dropped from the expression 
because they do not contribute to the integral in \eqref{G+_general} 
for a sufficiently large $\abs{\eta_0}$\,. 
We thus find that the Wightman function has the form 
\begin{align}
 G_{X,k}^+(\eta,\eta';\eta_0) 
 = G^+_{{\rm BD},k}(\eta,\eta') 
  + \Delta G_{k}^+(\eta,\eta';\eta_0)
\label{eq:G+expansion}
\end{align}
with%
\footnote{
Terms of order $(-k\eta_0)^{-3}$ also disappear 
since they necessarily include the rapidly oscillating factors. }
\begin{align}
 \Delta G_{k}^+(\eta,\eta';\eta_0)
 &= [(-\eta)(-\eta')]^{\frac{d-1}{2}}\, f(-k\,\eta_0)\, 
\nn\\
 &\quad\times
 \bigl[H^{(1)}_\nu(-e^{\ii\varepsilon}\,k\,\eta)\,
 H^{(2)}_\nu(-e^{-\ii\varepsilon}\,k\,\eta') 
\nn\\
 &\qquad
 +H^{(2)}_\nu(-e^{-\ii\varepsilon}\,k\,\eta)\,
  H^{(1)}_\nu(-e^{\ii\varepsilon}\,k\,\eta')\bigr] 
 +\mathcal{O}\bigl( (-k\,\eta_0)^{-4}\bigr)\,,
\label{DeltaG_k}
\\
 f(z)&\equiv
 \frac{\pi}{4}\Bigl(\frac{d-2}{4}\Bigr)^2\,z^{-2} \,.
\label{f-alpha}
\end{align}
Here, an infinitesimal positive constant $\varepsilon$ is introduced 
in order to make the $k$ integration in Eq.~\eqref{G+_general} finite 
(see footnote \ref{fn:regularization} in appendix \ref{app:Wightman-function}).
Then, comparing \eqref{DeltaG_k} and \eqref{f-alpha} with \eqref{f_ab-alpha}, 
we find that $\alpha = 2$ 
and thus the relaxation time is $\ell/\alpha = \ell/2$\,. 
Note that the reason why the relaxation time is given by $\ell/2$ (not by $\ell$)
is the disappearance of the order $(-k\,\eta_0)^{-1}$ terms from \eqref{Wightman_k_asympt} 
due to highly oscillatory integrals.

We close this subsection 
with a comment that the state $\ket{0_{\eta_0}}$ does not have a de Sitter invariance 
since it introduces an extra scale $\eta_0$ as a cutoff 
[even though the corresponding Wightman function 
has the same short distance behavior 
as $G_{\rm BD}^+(x,x')$]. 
This can be seen from the fact that $G^+(x,x';\eta_0)$ 
does not have a de Sitter invariant form 
(see appendix \ref{app:Wightman-function}).

\subsection{Higher-order corrections in $\dot{\cF}$}
\label{subsec:dS_cFdot}

With the Wightman function \eqref{eq:G+expansion}, 
$\dot{\cF}$ defined in \eqref{fdot_P} takes the form 
\begin{align} 
 \dot{\cF}(\Delta E,\tau,\tau_1;\,\eta_0)
 &=\dot{\cF}_{\rm BD}(\Delta E;\tau,\tau_1)
   +\Delta\dot{\cF}(\Delta E;\tau,\tau_1;\,\eta_0)\,,
\label{eq:cF-expansion}
\end{align}
where $\dot{\cF}_{\rm BD}(\Delta E;\tau,\tau_1)$ 
and $\Delta\dot{\cF}(\Delta E;\tau,\tau_1;\,\eta_0)$ 
are given by the integrals \eqref{fdot_BD_int} and \eqref{delta_fdot_int}, 
respectively.
The integrals 
can again be performed analytically as is done in appendix \ref{app:integration}, 
and we obtain 
[see \eqref{BD_fdot_asympt} and \eqref{delta_fdot_asympt} 
and recall that $\alpha=2$]
\begin{align}
 \dot{\cF}_{\rm BD}(\Delta E,\tau,\tau_1) 
 &\sim \dot{\cF}^{\eq} (\Delta E) 
 + {\rm const.}\, e^{-(\frac{d-1}{2}\pm\nu\pm\ii\Delta E)\,(\tau-\tau_1)} \,,
\label{eq:cF-0-tau1}
\\ 
 \Delta\dot{\cF}(\Delta E,\tau,\tau_1;\,\eta_0)
 &\sim e^{-2(\tau-\tau_0)}\,\Delta\dot{\cF}^{(0)} (\Delta E)
 +{\rm const.}\, e^{-(\frac{d-1}{2}\pm\nu\pm\ii\Delta E)\,(\tau-\tau_1)} \,,
\label{eq:cF-2-tau1}
\end{align}
where $\tau_0\equiv -\log(-\eta_0)$ 
and the coefficient of the leading term 
in $\Delta\dot{\cF}(\Delta E,\tau,\tau_1;\,\eta_0)$ is given by%
\footnote{
$ _3\widehat{F}_2$ is defined from 
the generalized hypergeometric function $_3F_2$ 
(see appendix \ref{app:G-function}) as 
\begin{align}
 \, _3\widehat{F}_2\Bigl(
  \afrac{a_1,\, a_2,\, a_3}
                     {b_1,\, b_2};\, z\Bigr) \equiv 
 \frac{\Gamma(a_1)\,\Gamma(a_2)\,\Gamma(a_3)}{\Gamma(b_1)\,\Gamma(b_2)}\,\, _3F_2\Bigl(
  \afrac{a_1,\, a_2,\, a_3}
                     {b_1,\, b_2};\, z\Bigr) \,.
\nn
\end{align}
} 
\begin{align}  
\Delta\dot{\cF}^{(0)} (\Delta E)&=\Bigl(\frac{d-2}{4}\Bigr)^2
\frac{\ii}{2^{5}\pi^{\frac{d-1}{2}}\Gamma(\frac{d-1}{2})}
\nn\\
&\quad\times\Biggl\{
 e^{\ii\pi\,\frac{d-2}{2}}\,\Biggl[
  \frac{e^{\ii \pi\,\nu}}{\sin (\pi\,\nu)}
     \, _3\widehat{F}_2\Biggl(
         \afrac{\frac{d-3}{2},\,
                             \frac{d-3}{2}+\nu,\,
                             \frac{\frac{d-1}{2}+\nu-2-\ii\Delta E}{2}}
                            {1+\nu,\,
                             \frac{\frac{d+3}{2}+\nu-2-\ii\Delta E}{2}};\,
                             e^{-\ii0}\Biggr)
\nn\\
  &\qquad\qquad\qquad + \frac{e^{-\ii\pi\,\nu}}{\sin (-\pi\,\nu)}
     \, _3\widehat{F}_2\Biggl(
         \afrac{\frac{d-3}{2},\,
                             \frac{d-3}{2}-\nu,\,
                             \frac{\frac{d-1}{2}-\nu-2-\ii\Delta E}{2}}
                            {1-\nu,\,
                             \frac{\frac{d+3}{2}-\nu-2-\ii\Delta E}{2}};\,
                             e^{-\ii0}\Biggr) \Biggr]
\nn\\
 &\quad\quad\quad +e^{-\ii\pi\,\frac{d-2}{2}}\,\Biggl[
  \frac{e^{-\ii \pi\,\nu}}{\sin (\pi\,\nu)}
     \, _3\widehat{F}_2\Biggl(
         \afrac{\frac{d-3}{2},\,
                             \frac{d-3}{2}+\nu,\,
                             \frac{\frac{d-1}{2}+\nu-2-\ii\Delta E}{2}}
                            {1+\nu,\,
                             \frac{\frac{d+3}{2}+\nu-2-\ii\Delta E}{2}};\,
                             e^{\ii0}\Biggr)
\nn\\ 
  &\qquad\qquad\qquad + \frac{e^{\ii\pi\,\nu}}{\sin (-\pi\,\nu)}
     \, _3\widehat{F}_2\Biggl(
         \afrac{\frac{d-3}{2},\,
                             \frac{d-3}{2}-\nu,\,
                             \frac{\frac{d-1}{2}-\nu-2-\ii\Delta E}{2}}
                            {1-\nu,\,
                             \frac{\frac{d+3}{2}-\nu-2-\ii\Delta E}{2}};\,
                             e^{\ii0}\Biggr) \Biggr]
\Biggr\}
\nn\\ 
&\quad\quad\quad
+(\Delta E \to -\Delta E)\,.
\label{eq:Fdot2}
\end{align}

\subsection{Thermalization of a two-level detector in de Sitter space}
\label{sec:thermal}

As we discussed in the preceding subsections, 
if we take the initial state for the scalar field 
to be the instantaneous ground state at a finite past, 
there exists a damping term in $\dot{\cF}$ with the relaxation time $\ell/2$. 
This behavior of $\dot{\cF}$ is expected 
to represent the relaxation of the surrounding medium.
In this subsection, as an analytically tractable example, 
we consider the case where the detector is a two-level system 
with energy eigenvalues $E_1$ and $E_2$ with $\Delta E \equiv E_2-E_1>0$\,, 
and describe how the density distribution of the detector 
approaches the Gibbs distribution with relaxation time $\ell/2$\,.

To proceed the analysis, we set the following assumptions:
\makeatletter
\renewcommand{\labelenumi}{(\roman{enumi})}
\renewcommand{\theenumi}{(\roman{enumi})}
\makeatother
\begin{enumerate} 
\item
The initial distribution of the detector, $\rho(\tau_1)$, 
is averaged over the initial proper time $\tau_1$ 
for the duration $\Delta\tau\gtrsim 1/\Delta E$\,. 
This is for describing distributions at different energy levels to a good accuracy. 
As a consequence, the off-diagonal elements of $\rho(\tau_1)$ 
can be effectively set to zero 
because they are oscillatory during the time. 
\label{enum:ave}
\item 
The energy difference should be much larger than the natural energy scale 
of de Sitter space, $\Delta E \gg \ell^{-1} = 1$\,,
in order for relaxation modes with relaxation times of order $\ell$ 
to be observed. 
\label{enum:DeltaE}
\item The operator $\mu$ is off-diagonal with respect to 
the basis $\ket{m}$ $(m=1,2)$, 
\begin{align}
 \mu = \bigl( \bra{m}\mu\ket{n} \bigr)
  =\begin{pmatrix}
    0 & \mu_{12} \cr
    \mu_{12}^\ast & 0
   \end{pmatrix}\,.
\end{align}
This is simply for making the following analysis much easier.
\label{enum:mu}
\end{enumerate}
\makeatletter
\renewcommand{\labelenumi}{\arabic}
\renewcommand{\theenumi}{\arabic}
\makeatother

One can easily show that the master equation \eqref{dense} decomposes 
into the diagonal and off-diagonal parts 
under assumption \ref{enum:mu}.  
The off-diagonal part is given by linear differential equations 
of the form 
\begin{align} 
 \begin{pmatrix}
  \dot{\rho}_{12}(\tau) \cr \dot{\rho}_{21}(\tau)
 \end{pmatrix}
 &= \mathcal{M}(\tau,\tau_1)\,
   \begin{pmatrix}
    \rho_{12}(\tau) \cr \rho_{21}(\tau)
   \end{pmatrix}\,,
\\ 
 \mathcal{M}(\tau,\tau_1)
 &\equiv
 \begin{pmatrix}
  \ii\Delta E-\abs{\mu_{12}}^2\,\dot{\cF}(0;\,\tau,\,\tau_1;\,\eta_0) &
  \mu_{12}^2\,\dot{\cF}(0;\,\tau,\,\tau_1;\,\eta_0) \cr
  \mu_{12}^{\ast\,2}\,\dot{\cF}(0;\,\tau,\,\tau_1;\,\eta_0) &
  -\,\ii \Delta E- \abs{\mu_{12}}^2\,\dot{\cF}(0;\,\tau,\,\tau_1;\,\eta_0)
 \end{pmatrix}\,,
\end{align}
which can be integrated to
\begin{align}
 \begin{pmatrix}
  \rho_{12}(\tau) \cr \rho_{21}(\tau)
 \end{pmatrix}
 ={\rm T}\exp\Bigl(\int_{\tau_1}^\tau\!\!\rmd \tau'\,\mathcal{M}(\tau',\tau_1)\Bigr)
  \begin{pmatrix}
  \rho_{12}(\tau_1) \cr \rho_{21}(\tau_1)
 \end{pmatrix}. 
\end{align}
Since we can effectively set $\rho_{12}(\tau_1)=\rho_{21}(\tau_1)=0$ 
due to assumption \ref{enum:ave}, 
we can also set $\rho_{12}(\tau)=\rho_{21}(\tau)=0$ 
[assuming that $\mathcal{M}(\tau',\tau_1)$ changes slowly 
when averaging over $\tau_1$].

The diagonal part can then be obtained from \eqref{master} and \eqref{wmk} as
\begin{align}
 \begin{pmatrix}
  \dot{\rho}_{11}(\tau) \cr \dot{\rho}_{22}(\tau)
 \end{pmatrix}
 =
 \begin{pmatrix}
  -w_{21}(\tau,\tau_1;\eta_0) & w_{12}(\tau,\tau_1;\eta_0) \cr
  w_{21}(\tau,\tau_1;\eta_0) & -w_{12}(\tau,\tau_1;\eta_0)
 \end{pmatrix}
 \begin{pmatrix}
  \rho_{11}(\tau) \cr \rho_{22}(\tau)
 \end{pmatrix}
\end{align}
with
\begin{align}
 w_{m k}(\tau,\tau_1;\eta_0)=\lambda^2\,\abs{\mu_{mk}}^2\,
 \dot\cF(E_m-E_k; \tau,\tau_1; \eta_0)\,.
\label{wmk_reduced}
\end{align}
Using Eqs.~\eqref{eq:cF-expansion}, \eqref{eq:cF-0-tau1} and \eqref{eq:cF-2-tau1} 
together with assumptions \ref{enum:ave} 
and \ref{enum:DeltaE}, 
one can easily show that $\dot{\cF}$ comes to take the following form
after averaging over $\tau_1$\,:
\begin{align} 
 \dot{\cF}_{\rm av}(\Delta E;\tau;\eta_0)
 &\equiv \frac{1}{\Delta\tau}\int^{\tau_1+\Delta\tau}_{\tau_1}\!\!\rmd \tau^{\prime}_1\,
 \dot{\cF}(\Delta E;\tau,\tau^{\prime}_1;\eta_0)
\nn\\
 &=\dot{\cF}^{\eq}(\Delta E) +e^{-2(\tau-\tau_0)}\,\Delta\dot{\cF}^{(0)}(\Delta E)
 + \mathcal{O}(e^{-4(\tau-\tau_0)})\,,
\label{eq:average}
\end{align}
where $\dot{\cF}^{\eq}(\Delta E)$ and $\Delta\dot{\cF}^{(0)}(\Delta E)$ are 
given by \eqref{eq:Fdot-eq} and \eqref{eq:Fdot2}. 
Note that the uninteresting $\tau_1$-dependent terms 
proportional to $e^{-(\frac{d-1}{2}\pm\nu\pm\ii\Delta E)\,(\tau-\tau_1)}$ 
have totally disappeared from $\dot{\cF}$ due to the averaging procedure 
[see \eqref{eq:cF-0-tau1} and \eqref{eq:cF-2-tau1}].
Replacing $\dot\cF(\Delta E; \tau,\tau_1;\eta_0)$ in \eqref{wmk_reduced}
by $\dot\cF_{\rm av}(\Delta E; \tau;\eta_0)$, 
we obtain the master equation for the diagonal elements of the form 
\begin{align}
\begin{pmatrix}
 \dot{\rho}_{11}(\tau) \cr
 \dot{\rho}_{22}(\tau)
 \end{pmatrix}
&=
\begin{pmatrix}
 -w_{+}(\tau;\tau_0) &w_{-}(\tau;\tau_0)\cr
 w_{+}(\tau;\tau_0) &-w_{-}(\tau;\tau_0)
\end{pmatrix}
\begin{pmatrix}
 \rho_{11}(\tau) \cr
 \rho_{22}(\tau)
\end{pmatrix} 
\label{2level-master}
\end{align}
with
\begin{align}
 w_{\pm}(\tau;\tau_0) &\equiv
 \lambda^2\,\abs{\mu_{12}}^2\,\bigl[
 \dot{\cF}^{\eq}(\pm\Delta E) +e^{-2(\tau-\tau_0)}\,\Delta\dot{\cF}^{(0)}(\pm\Delta E)
 +{\cal O}\bigl(e^{-4(\tau-\tau_0)}\bigr)
 \bigr]\,.
\end{align}

From a general argument following \eqref{eq:Fdot-eq},
we have already seen that, as $\tau$ becomes large,  
the density distribution $\rho_{mm}(\tau)$ approaches the Gibbs distribution 
at temperature $1/2\pi\ell$\,,
\begin{align}
 \rho^{\eq}_{mm}
 = \frac{e^{-2\pi\ell E_m}}{Z} \quad
 \bigl(Z=e^{-2\pi\ell E_1}+e^{-2\pi\ell E_2}\bigr)\,. 
\end{align}
In order to investigate how the detector relaxes to this equilibrium, 
we expand $\rho_{mm}(\tau)$ as 
$\rho_{mm}(\tau)\equiv \rho^{\eq}_{mm}+\Delta\rho_{mm}(\tau)$ 
and keep small quantities to the first order, 
assuming that $\Delta\rho_{mm}/\rho^{\eq}_{mm}\ll 1$   
and $e^{-2(\tau-\tau_0)}\ll 1$\,. 
Then \eqref{2level-master} becomes
\begin{align}
 \frac{\rmd}{\rmd\tau}
 \Delta\rho_{11}(\tau)
 &=-\lambda^2\,\abs{\mu_{12}}^2\,\bigl[\,
 e^{-2(\tau-\tau_0)}\,\Delta\dot{\cF}^{(0)}(\Delta E)\,\tanh(\pi\Delta E)
\nn\\
 &\qquad\qquad\qquad  +\bigl(\dot{\cF}^{\eq}(\Delta E)+\dot{\cF}^{\eq}(-\Delta E)\bigr)\,\Delta\rho_{11}(\tau)\,
\bigr]\,.
\label{master11}
\end{align}
We used the fact that
$\Delta\dot{\cF}^{(0)}(\Delta E)$ is an even function of $\Delta E$ 
[see \eqref{eq:Fdot2}] 
and $\Delta\rho_{11}=-\Delta\rho_{22}$\,.  
This equation can be solved easily, 
and we find that the relaxation behavior of the density distribution 
is given by 
\begin{align} 
 &\Delta\rho_{11}(\tau)
\nn\\ 
 &=e^{-\lambda^2\,\abs{\mu_{12}}^2\,A(\Delta E)\,(\tau-\tau_2)}
 \Delta\rho_{11}(\tau_2)
\nn\\ 
 &\quad-\frac{\lambda^2\,\abs{\mu_{12}}^2\,\Delta\dot{\cF}^{(0)}(\Delta E)}
         {\lambda^2\,\abs{\mu_{12}}^2\,A(\Delta E)-2}
 \,\tanh(\pi\Delta E)
 \bigl( e^{-2(\tau-\tau_2)}
 -e^{-\lambda^2\,\abs{\mu_{12}}^2\,A(\Delta E)\,(\tau-\tau_2)-2(\tau_2-\tau_0)}
 \bigr) \,.
\label{relax}
\end{align}
Here,
\begin{align}
 A(\Delta E)\equiv \dot{\cF}^{\rm eq}(\Delta E)+\dot{\cF}^{\rm eq}(-\Delta E) \,,
\label{eq:A-def}
\end{align}
and $\tau_2$ is an arbitrary proper time after $\tau_1$ $(\tau_2>\tau_1)$, 
where the linear approximation is well justified.

We note that the function $A(\Delta E)$ 
has the following asymptotic form for large $\abs{\Delta E}$\,:
\begin{align}
 A(\Delta E) \sim \frac{\abs{\Delta E}^{d-3}}
                       {2^{d-2}\,\pi^{\frac{d-3}{2}}\,\Gamma\bigl(\frac{d-1}{2}\bigr)} 
 \qquad \bigl(\,\abs{\Delta E}\to \infty\bigr)\,,
\label{eq:A-DeltaE}
\end{align}
as can be easily shown 
by using the asymptotic form of the Gamma function (8.328-1 of \cite{GR}),
\begin{align}
 \lim_{\abs{y}\to \infty} \abs{\Gamma(x+\ii y)}
 \sim \sqrt{2\pi}\,e^{-\frac{\pi}{2}\,\abs{y}}\, 
 \abs{y}^{x-\frac{1}{2}}\quad (x,\,y\in\mathbb{R})\,.
\end{align}
We now consider a detector that satisfies the inequality 
\begin{align}
 \lambda^2\,\abs{\mu_{12}}^2\,(\Delta E)^{d-3}\,\ell\gg 1 \,.
\label{eq:ideal-assumption}
\end{align}
Then, by using the asymptotic form \eqref{eq:A-DeltaE}, 
$A(\Delta E)\sim \abs{\Delta E}^{d-3}$\,,
and the inequality \eqref{eq:ideal-assumption}, 
the damping terms in \eqref{relax} 
which are proportional to 
$e^{-\lambda^2\,\abs{\mu_{12}}^2\,A(\Delta E)\,(\tau-\tau_2)}$ 
rapidly disappear from the expression, 
leaving only the term proportional to $e^{{}-2(\tau-\tau_2)}$\,, 
\begin{align}
 \Delta\rho_{11}(\tau) \sim {}
 -\frac{\Delta\dot{\cF}^{(0)}(\Delta E)}
         {A(\Delta E)}
 \,\tanh(\pi\ell\Delta E)\,e^{-2(\tau-\tau_2)/\ell}\,,
\label{eq:universal}
\end{align}
where we have restored the curvature radius $\ell$\,. 
Note that the coefficient $\Delta\dot{\cF}^{(0)}(\Delta E)/A(\Delta E)$ 
does not depend on details of the detector 
(such as $\lambda \,\mu_{12}$).
The remaining damping term in \eqref{eq:universal} 
corresponds to the desired relaxation mode 
with the relaxation time $\ell/2$\,. 
Since only those rapidly disappearing terms depend on details of detector, 
one may say that such detector under consideration is an {\em ideal detector}, 
in the sense that 
it quickly loses its own nonequilibrium properties 
and gets adjusted to its environment almost instantaneously.

\section{Conclusion and discussions}
\label{sec:conclusion}

In this paper, we have considered an Unruh-DeWitt detector 
staying in the Poincar\'e patch of de Sitter space. 
The main difference of our setup from those in the literature 
is in that the scalar field 
(before interacting with the detector) 
is not in the Bunch-Davies vacuum (nor in the $\alpha$-vacuum). 
Then the Unruh-DeWitt detector behaves 
as if it is in a nonequilibrium environment. 
In order to deal with such situations, 
we first derived the master equation 
which describes a finite time evolution of 
the density matrix of an Unruh-DeWitt detector in arbitrary geometry. 

We then applied the framework to de Sitter space. 
We showed that there exists a damping term in $\dot{\cF}$ 
with a relaxation time of the form $\ell/\alpha$ 
if the initial state of scalar field 
is chosen such that the Wightman function 
takes the form \eqref{Wightman-decomposition} with \eqref{eq:DeltaG+} and \eqref{f_ab-alpha}. 
In particular, 
if we take the initial state to be the instantaneous ground state at a finite past,
the relaxation time is always given by $\ell/2$\,. 
We further gave an explicit description of the relaxation process 
for a two-level detector.

We here should stress again that we are not considering just 
the thermalization process of a detector dipped in a thermal bath. 
In fact, since the initial state is chosen 
to be different from the Bunch-Davies vacuum, 
the detector should initially behave as if it is in a medium 
which is not in thermodynamic equilibrium.  
As time goes on, the detector comes to behave as if it is in a thermal bath, 
since the difference of the initial state from the Bunch-Davies vacuum 
becomes irrelevant at later times 
(i.e., $\Delta\dot{\cF}/\dot{\cF}_{\rm BD} \ll 1$ at later times).

In reality, there can be many relaxation processes for such a detector, 
some of which are simply the processes 
where the detector gets adjusted to its environment. 
However, these processes usually depend on details of the detector, 
and thus can be neglected 
by considering an ideal detector 
which quickly responds to changes in its environment. 
The terms proportional to 
$e^{-\lambda^2\,\abs{\mu_{12}}^2\,A(\Delta E)\,\tau}$ 
in \eqref{relax} actually represent such processes. 
On the other hand, there are terms including the factors 
$e^{-(\frac{d-1}{2}\pm{\rm Re}\,\nu)\,\tau}$ in $\dot{\cF}$. 
This kind of terms always exist 
even when the initial state is the Bunch-Davies vacuum, 
and thus do not have relevance to the relaxation of the nonequilibrium medium 
depicted in Fig.~\ref{fig:medium} (b). 
For the case of two-level detector considered in section \ref{sec:thermal}, 
such damping terms disappear from the expression 
after taking an average over the start-up time $\tau_1$\,.  
There is also a damping term proportional to $e^{-\alpha\,\tau/\ell}$ 
[see \eqref{delta_fdot_asympt}], 
which appears only when the initial state is different from the Bunch-Davies vacuum 
and thus describes the nonequilibrium dynamics of the surrounding medium. 
In particular, the relaxation time takes a universal value $\ell/2$ 
(i.e.\ $\alpha=2$) 
for the instantaneous ground states at finite pasts. 
We expect that the relaxation time $\ell/2$ 
gives a quantity representing the nonequilibrium thermodynamic character 
intrinsic to de Sitter space, 
just as the temperature of the final Gibbs distribution 
represents the Gibbons-Hawking temperature intrinsic to de Sitter space.

In the previous studies,
thermal properties of de Sitter space
have been investigated mainly in the context of equilibrium thermodynamics. 
The relaxation processes discussed in this paper 
may serve as examples related to the nonequilibrium dynamics 
of de Sitter space.

\section*{Acknowledgments}

We thank Y.~Hamada, H.~Ishimori, T.~Kameyama, H.~Kawai, 
K.~Murase, K.~Oda and A.~Ogasahara 
for useful discussions on the Unruh-DeWitt detector. 
This work was supported by the Grant-in-Aid for the Global COE program 
``The Next Generation of Physics, Spun from Universality and
Emergence" from the Ministry of Education, Culture, Sports, 
Science and Technology (MEXT) of Japan. 
This work was also supported by MEXT (Grant No.\,23540304).

\appendix

\section{Derivation of the master equation}
\label{app:master}
In this appendix, 
we derive the master equation \eqref{master_full}.  
Since $\rho^{\rm tot}_I(t)$ satisfies the time evolution equation \eqref{vn},
$\PP\rho^{\rm tot}_I(t)$ and $\QQ\rho^{\rm tot}_I(t)$ 
satisfy the differential equations 
(note that $\PP+\QQ=1$)
\begin{align}
 \frac{\rmd}{\rmd t}\PP\rho^{\rm tot}_I(t)
 &=-\ii\PP\ad_{V_I(t)} \PP\rho^{\rm tot}_I(t)
    -\ii\PP\ad_{V_I(t)} \QQ\,\rho^{\rm tot}_I(t)\,,
\label{vnp}
\\
 \frac{\rmd}{\rmd t}\QQ\rho^{\rm tot}_I(t)
 &=-\ii\QQ\ad_{V_I(t)} \PP\rho^{\rm tot}_I(t)
   -\ii\QQ\ad_{V_I(t)} \QQ\,\rho^{\rm tot}_I(t)\,.
\label{vnq}
\end{align}
The solution of \eqref{vnq} is given by 
\begin{align} 
 &\QQ \rho^{\rm tot}_I(t)
\nn\\ 
 &=-\ii\int^{t}_{t_1}\rmd t'\,
 {\rm T} e^{-\ii\int^{t}_{t'}\rmd t^{\prime\prime}
 \QQ\ad_{V_I(t^{\prime\prime})}}
 \QQ\ad_{V_I(t')}\,\PP\rho^{\rm tot}_I(t')
 + {\rm T}e^{-\ii\int^{t}_{t_1}\rmd t^{\prime}\QQ
 \ad_{V_I(t^{\prime})}}\,\QQ\rho^{\rm tot}_I(t_1)\,,
\label{solutionQ}
\end{align}
and substituting this to \eqref{vnp} 
and using $(\rmd /\rmd t)\PP\rho^{\rm tot}_I(t)
=(\rmd\rho_I(t)/\rmd t)\otimes X^\phi$\,,
we obtain
\begin{align} 
 \frac{\rmd \rho_I(t)}{\rmd t} \otimes X^\phi=
 &-\ii\PP\ad_{V_I(t)} \PP\rho^{\rm tot}_I(t)
 -\ii\PP\ad_{V_I(t)}
 {\rm T}e^{-\ii\int^{t}_{t_1}\rmd t^{\prime}\QQ\ad_{V_I(t^{\prime})}}
 \QQ\,\rho^{\rm tot}_I(t_1)
\nn\\ 
 &-\PP\ad_{V_I(t)}\int^{t}_{t_1}\rmd t'\,
{\rm T} e^{-\ii\int^{t}_{t'}\rmd t^{\prime\prime}
 \QQ \ad_{V_I(t^{\prime\prime})}}
 \QQ\,\ad_{V_I(t')} \PP\rho^{\rm tot}_I(t')\,.
\label{pdenseevo}
\end{align}
This equation can be simplified 
by noticing that  $X^\phi$ can be chosen arbitrarily 
without changing the time evolution of $\rho_I(t)$\,. 
If we set $X^\phi=\rho^\phi(t_1)=\rho^\phi_I(t_1)$\,, 
we have Eq.~\eqref{pq_simple}: 
$\PP\rho^{\rm tot}_I(t_1)=\rho^{\rm tot}_I(t_1)$\,,
$\QQ\rho^{\rm tot}_I(t_1)=0$\,. 
Thus, 
the second term on the right-hand side of \eqref{pdenseevo}
vanishes.
Furthermore, 
since the interaction has the factorized form 
$V_I(t)= \lambda\,\frac{\rmd \tau}{\rmd t}\,\mu_I(\tau) \otimes 
\phi_I\bigl(x(\tau)\bigr)\,\theta(t-t_1)$ 
[Eq.~\eqref{vint}],
we obtain
\begin{align}
 \Tr_{\phi} \bigl(\ad_{V_I(t)} \PP\rho^{\rm tot}_I(t) \bigr)
 = \lambda\,\frac{\rmd \tau}{\rmd t}\,\bigl[\mu_I(\tau), \,\rho_I(t)\bigr]
 \Tr_\phi \bigl( \phi_I\bigl(x(\tau)\bigr) \rho_I^\phi(t_1) \bigr) \,,
\end{align}
which vanishes when the condition \eqref{1pt0} holds.
We thus find that the first term in \eqref{pdenseevo} also vanishes, 
\begin{align}
 \PP\ad_{V_I(t)} \PP\rho^{\rm tot}_I(t)
  =\Tr_{\phi} \bigl(\ad_{V_I(t)} \PP\rho^{\rm tot}_I(t) \bigr)\otimes X^\phi
  =0\,.
\label{first_vanish}
\end{align}
The last equation \eqref{first_vanish} shows that
$
 \QQ\ad_{V_I(t)} \PP\rho^{\rm tot}_I(t)=\ad_{V_I(t)} \PP\rho^{\rm tot}_I(t)
  =\ad_{V_I(t)}\,\bigl( \rho_I(t) \otimes \rho^\phi_I(t_1) \bigr)\,,
$
and thus, \eqref{pdenseevo} becomes the master equation \eqref{master_full}.

\section{Meijer's $G$-function and generalized hypergeometric function}
\label{app:G-function}

Meijer's $G$-function $G^{m,n}_{p,q}$ ($0\leq m\leq q$, $0\leq n\leq p$) 
is defined by 
(see 9.302 of \cite{GR} for details on the choice of the contour $C$)
\begin{align} 
 G^{m,n}_{p,q}\Bigl(z\,\Big\vert
  \afrac{a_1,...,a_p}{b_1,...,b_q}\Bigr)
 = \int_C\frac{\rmd s}{2\pi\ii}\,
  \frac{\prod_{j=1}^m\Gamma(b_j-s)\,\prod_{j=1}^n\Gamma(1-a_j+s)}
       {\prod_{j=m+1}^q\Gamma(1-b_j+s)\,\prod_{j=n+1}^p\Gamma(a_j-s)}\,z^s\,.
\end{align}
This function is invariant under arbitrary permutation of a set 
$\{a_1,...,a_n\}$, $\{a_{n+1},...,a_p\}$, 
$\{b_1,...,b_m\}$, or $\{b_{m+1},...,b_q\}$. 
As a special case of the $G$-function, 
the generalized hypergeometric function is defined by (9.34-8 of \cite{GR})
\begin{align} 
 \, _pF_q\Bigl(
    \afrac{a_1,...,a_p}
                       {b_1,...,b_q};\,z\Bigr) 
 &\equiv \frac{\prod_{i=1}^q\Gamma(b_i)}
        {\prod_{i=1}^p\Gamma(a_i)}\,
   G^{1,p}_{p,q+1}\Bigl(-z\,\Big\vert
    \afrac{1-a_1,...,1-a_p}{0,1-b_1,...,1-b_q}\Bigr)
\nn\\
 &= \frac{\prod_{i=1}^q\Gamma(b_i)}
         {\prod_{i=1}^p\Gamma(a_i)}\,
  \int_C\frac{\rmd s}{2\pi\ii}\,
  \frac{\Gamma(-s)\,\prod_{i=1}^p\Gamma(a_i+s)}
       {\prod_{i=1}^q\Gamma(b_i+s)}\,(-z)^s \,.
\label{eq:pFq-def}
\end{align}
For convenience, 
we define the following functions using the generalized hypergeometric functions:
\begin{align}
 \, _p\widehat{F}_q\Bigl(
    \afrac{a_1,...,a_p}
                       {b_1,...,b_q};\,z\Bigr)
 \equiv
       \frac{\prod_{i=1}^p\Gamma(a_i)}
            {\prod_{i=1}^q\Gamma(b_i)}\,
 \, _pF_q\Bigl(
    \afrac{a_1,...,a_p}
                       {b_1,...,b_q};\,z\Bigr) \,.
\end{align}

If no two $b_i$ ($1\leq i\leq m$) differ by an integer, 
Meijer's $G$-function $G^{m,n}_{p,q}$ with $p<q$, 
or $p=q$ and $m+n>p$, or $p=q$ and $m+n=p$ and $\abs{z}<1$, 
can be expanded by using the generalized hypergeometric functions \cite{Wolfram},
\begin{align} 
 G^{m,n}_{p,q}\Bigl(z\,\Big\vert
  \afrac{a_1,...,a_p}{b_1,...,b_q}\Bigr)
 &= \sum_{k=1}^m\frac{\prod_{i=1,i\neq k}^m\Gamma(b_i-b_k)\,\prod_{i=1}^n\Gamma(1-a_i+b_k)}
       {\prod_{i=n+1}^p\Gamma(a_i-b_k)\,\prod_{i=m+1}^q\Gamma(1-b_i+b_k)}
    \,z^{b_k}
\nn\\ 
 &\quad\times \, _pF_{q-1}\Bigl(
          \afrac{1-a_1+b_k,...,
                              1-a_p+b_k}
                             {1-b_1+b_k,...,*,...,
                              1-b_q+b_k};\,(-1)^{p-m-n}z\Bigr) \,,
\end{align}
where $*$ means that the $k^{\rm th}$ term has been omitted. 
In particular, $G^{2,2}_{3,3}$ is expanded as 
\begin{align} 
 &G^{2,2}_{3,3}\Bigl(z\,\Big\vert
  \afrac{a_1,a_2,a_3}{b_1,b_2,b_3}\Bigr)
\nn\\ 
 &= z^{b_1}\,\frac{\Gamma(b_2-b_1)\,\Gamma(1-b_2+b_1)}{\Gamma(a_3-b_1)\,\Gamma(1-a_3+b_1)}\,
    \, _3\widehat{F}_2\Bigl(
          \afrac{1-a_1+b_1,\,
                              1-a_2+b_1,\,
                              1-a_3+b_1}
                             {1-b_2+b_1,\,
                              1-b_3+b_1};\,-z\Bigr) 
\nn\\ 
 &+ z^{b_2}\,\frac{\Gamma(b_1-b_2)\,\Gamma(1-b_1+b_2)}{\Gamma(a_3-b_2)\,\Gamma(1-a_3+b_2)}\,
    \, _3\widehat{F}_2\Bigl(
          \afrac{1-a_1+b_2,\,
                              1-a_2+b_2,\,
                              1-a_3+b_2}
                             {1-b_1+b_2,\,
                              1-b_3+b_2};\,-z\Bigr) \,.
\end{align}
Furthermore, one can show
\begin{align} 
 &G^{2,3}_{4,4}\Bigl(z\,\Big\vert
  \afrac{a_1,a_2,a_3,a_4}{b_1,b_2,b_3,a_4}\Bigr)
\nn\\ 
 &= z^{b_1}\,\frac{\Gamma(b_2-b_1)\,\Gamma(1-b_2+b_1)}{\Gamma(a_4-b_1)\,\Gamma(1-a_4+b_1)}\,
\nn\\ 
 &\qquad \times   \, _4\widehat{F}_3\Bigl(
          \afrac{1-a_1+b_1,\,
                              1-a_2+b_1,\,
                              1-a_3+b_1,\,
                              1-a_4+b_1}
                             {1-b_2+b_1,\,
                              1-b_3+b_1,\,
                              1-a_4+b_1};\,-z\Bigr) 
\nn\\ 
 &\quad+ z^{b_2}\,\frac{\Gamma(b_1-b_2)\,\Gamma(1-b_1+b_2)}{\Gamma(a_4-b_2)\,\Gamma(1-a_4+b_2)}\,
\nn\\ 
 &\qquad\times    \, _4\widehat{F}_3\Bigl(
          \afrac{1-a_1+b_2,\,
                              1-a_2+b_2,\,
                              1-a_3+b_2,\,
                              1-a_4+b_2}
                             {1-b_1+b_2,\,
                              1-b_3+b_2,\,
                              1-a_4+b_2};\,-z\Bigr) 
\nn\\ 
 &= z^{b_1}\,\frac{\Gamma(b_2-b_1)\,\Gamma(1-b_2+b_1)}{\Gamma(a_4-b_1)\,\Gamma(1-a_4+b_1)}\,
    \, _3\widehat{F}_2\Bigl(
          \afrac{1-a_1+b_1,\,
                              1-a_2+b_1,\,
                              1-a_3+b_1}
                             {1-b_2+b_1,\,
                              1-b_3+b_1};\,-z\Bigr) 
\nn\\ 
 &\quad+ z^{b_2}\,\frac{\Gamma(b_1-b_2)\,\Gamma(1-b_1+b_2)}{\Gamma(a_4-b_2)\,\Gamma(1-a_4+b_2)}\,
    \, _3\widehat{F}_2\Bigl(
          \afrac{1-a_1+b_2,\,
                              1-a_2+b_2,\,
                              1-a_3+b_2}
                             {1-b_1+b_2,\,
                              1-b_3+b_2};\,-z\Bigr)  \,,
\end{align}
by using the following identity 
[which can be derived from \eqref{eq:pFq-def}]: 
\begin{align}
 \, _4F_3\Bigl(\afrac{a_1,a_2,a_3,a_4}{b_1,b_2,a_4};\,z\Bigr)
 = \, _3F_2\Bigl(\afrac{a_1,a_2,a_3}{b_1,b_2};\,z\Bigr)\,.
\end{align}
From the above identities, one can show the following relations
for $\sigma,\,\sigma' =\pm 1$ and $\alpha\in\bRR$: 
\begin{align} 
 &G_{3,3}^{2,2}\Biggl(-e^{-(\sigma-\sigma')\,\ii 0}\,x\Biggl\vert
         \afrac{-\frac{d-\alpha-3}{2},\,
                             \frac{\alpha-\frac{d-5}{2}+\nu \pm \ii\Delta E}{2},\,
                             \nu-\frac{d-\alpha-3}{2}}
                            {\nu,\,
                             0,\,
                             \frac{\alpha-\frac{d-1}{2}+\nu \pm \ii\Delta E}{2}}\Biggr) 
\nn\\ 
 &=-\frac{\bigl(-e^{-(\sigma-\sigma')\,\ii 0}\,x\bigr)^{\nu}\, \sin \bigl(\pi\frac{d-1-\alpha}{2}\bigr)}{\sin (\pi\,\nu)}
     \, _3\widehat{F}_2\Biggl(
         \afrac{\frac{d-1-\alpha}{2},\,
                             \frac{d-1-\alpha}{2}+\nu,\,
                             \frac{\frac{d-1}{2}+\nu-\alpha \mp \ii\Delta E}{2}}
                            {1+\nu,\,
                             \frac{\frac{d+3}{2}+\nu-\alpha \mp \ii\Delta E}{2}};\,
                             e^{-(\sigma-\sigma')\,\ii0}\,x\Biggr) 
\nn\\ 
  &\quad -\frac{\sin \bigl(\pi\,\frac{d-1-\alpha-2\nu}{2}\bigr)}{\sin (-\pi\,\nu)}
     \, _3\widehat{F}_2\Biggl(
         \afrac{\frac{d-1-\alpha}{2},\,
                             \frac{d-1-\alpha}{2}-\nu,\,
                             \frac{\frac{d-1}{2}-\nu-\alpha \mp \ii\Delta E}{2}}
                            {1-\nu,\,
                             \frac{\frac{d+3}{2}-\nu-\alpha \mp \ii\Delta E}{2}};\,
                             e^{-(\sigma-\sigma')\,\ii0}\,x\Biggr) \,,
\\ 
 &G_{4,4}^{2,3}\Biggl(-e^{-(\sigma-\sigma')\,\ii 0}\,x\Biggl\vert
        \afrac{-\frac{d-\alpha-3}{2},\,
                            \frac{\alpha-\frac{d-5}{2}+\nu \pm \ii\Delta E}{2},\,
                            \nu-\frac{d-\alpha-3}{2},\,
                            \nu-\frac{d-\alpha-4}{2}}
                           {\nu,\,
                            0,\,
                            \frac{\alpha-\frac{d-1}{2}+\nu \pm \ii\Delta E}{2},\,
                            \nu-\frac{d-\alpha-4}{2}}\Biggr) 
\nn\\ 
 &= \frac{\bigl(-e^{-(\sigma-\sigma')\,\ii 0}\,x\bigr)^\nu\, \cos(\pi\,\frac{d-1-\alpha}{2})}{\sin(\pi\,\nu)}
   \, _3\widehat{F}_2\Biggl(
         \afrac{\frac{d-1-\alpha}{2},\,
                             \frac{d-1-\alpha}{2}+\nu,\,
                             \frac{\frac{d-1}{2}+\nu-\alpha \mp \ii\Delta E}{2}}
                            {1+\nu\,,
                             \frac{\frac{d+3}{2}+\nu-\alpha \mp \ii\Delta E}{2}};
                             e^{-(\sigma-\sigma')\,\ii0}\,x\Biggr) 
\nn\\ 
 &\quad + \frac{\cos\bigl(\pi\,\frac{d-1-\alpha-2\nu}{2}\bigr)}{\sin(-\pi\,\nu)} 
 \, _3\widehat{F}_2\Biggl(
         \afrac{\frac{d-1-\alpha}{2},\,
                             \frac{d-1-\alpha}{2}-\nu,\,
                             \frac{\frac{d-1}{2}-\nu-\alpha \mp \ii\Delta E}{2}}
                            {1-\nu\,,
                             \frac{\frac{d+3}{2}-\nu-\alpha \mp \ii\Delta E}{2}};
                             e^{-(\sigma-\sigma')\,\ii0}\,x\Biggr) \,,
\end{align}
from which we obtain the following formula for $x>0$:
\begin{align}
 &G_{3,3}^{2,2}\Biggl(-e^{-\ii(\sigma-\sigma')\,0}\,x\Biggl\vert
         \afrac{-\frac{d-\alpha-3}{2},\,
                             \frac{\alpha-\frac{d-5}{2} \pm \ii\Delta E+\nu}{2},\,
                             \nu-\frac{d-\alpha-3}{2}}
                            {\nu,\,
                             0,\,
                             \frac{\alpha-\frac{d-5}{2} \pm \ii\Delta E+\nu}{2}-1}\Biggr) 
\nn\\ 
 &+\ii\sigma\,G_{4,4}^{2,3}\Biggl(-e^{-\ii(\sigma-\sigma')\,0}\,x\Biggl\vert
        \afrac{-\frac{d-\alpha-3}{2},\,
                            \frac{\alpha-\frac{d-5}{2}+\nu \pm \ii\Delta E}{2},\,
                            \nu-\frac{d-\alpha-3}{2},\,
                           \nu-\frac{d-\alpha-4}{2}}
                           {\nu,\,
                            0,\,
                            \frac{\alpha-\frac{d-5}{2}+\nu \pm \ii\Delta E}{2}-1,\,
                            \nu-\frac{d-\alpha-4}{2}}\Biggr) 
\nn\\ 
 &= e^{\sigma\,\ii\pi\,\frac{d-\alpha}{2}}\,\Biggl[
  \frac{\bigl(-e^{-(\sigma-\sigma')\,\ii 0}\bigr)^\nu\,x^{\nu}}
       {\sin (\pi\,\nu)}
     \, _3\widehat{F}_2\Biggl(
         \afrac{\frac{d-1-\alpha}{2},\,
                             \frac{d-1-\alpha}{2}+\nu,\,
                             \frac{\frac{d-1}{2}+\nu-\alpha \mp \ii\Delta E}{2}}
                            {1+\nu,\,
                             \frac{\frac{d+3}{2}+\nu-\alpha \mp \ii\Delta E}{2}};\,
                             e^{-\ii(\sigma-\sigma')\,0}\,x\Biggr)
\nn\\ 
  &\qquad\qquad\qquad + \frac{e^{-\sigma\,\ii\pi\,\nu}}{\sin (-\pi\,\nu)}
     \, _3\widehat{F}_2\Biggl(
         \afrac{\frac{d-1-\alpha}{2},\,
                             \frac{d-1-\alpha}{2}-\nu,\,
                             \frac{\frac{d-1}{2}-\nu-\alpha \mp \ii\Delta E}{2}}
                            {1-\nu,\,
                             \frac{\frac{d+3}{2}-\nu-\alpha \mp \ii\Delta E}{2}};\,
                             e^{-\ii(\sigma-\sigma')\,0}\, x\Biggr) \Biggr]\,.
\label{eq:G+iG-3F2}
\end{align}

Another useful formula can be obtained
by using the identities \cite{Wolfram2}
\begin{align} 
 &\, _3\widehat{F}_2\Bigl(
          \afrac{a_1,\,a_2,\,a_3}{b_1,\,b_2};\,z
         \Bigr) 
\nn\\ 
 &= (-z)^{-a_1}\, \frac{\Gamma(a_1)\,\Gamma(a_2-a_1)\,\Gamma(a_3-a_1)}
                       {\Gamma(b_1-a_1)\,\Gamma(b_2-a_1)} \, _3F_2\Bigl(
          \afrac{a_1,\,a_1-b_1+1,\,a_1-b_2+1}
                             {a_1-a_2+1,\,a_1-a_3+1}\,;\,
             z^{-1}\Bigr) 
\nn\\ 
 &\quad +(-z)^{-a_2}\, \frac{\Gamma(a_2)\,\Gamma(a_1-a_2)\,\Gamma(a_3-a_2)}
                       {\Gamma(b_1-a_2)\,\Gamma(b_2-a_2)} \, _3F_2\Bigl(
          \afrac{a_2,\,a_2-b_1+1,\,a_2-b_2+1}
                             {a_2-a_1+1,\,a_2-a_3+1}\,;\,
             z^{-1}\Bigr) 
\nn\\ 
 &\quad +(-z)^{-a_3}\, \frac{\Gamma(a_3)\,\Gamma(a_1-a_3)\,\Gamma(a_2-a_3)}
                       {\Gamma(b_1-a_3)\,\Gamma(b_2-a_3)} \, _3F_2\Bigl(
          \afrac{a_3,\,a_3-b_1+1,\,a_3-b_2+1}
                             {a_3-a_1+1,\,a_3-a_2+1}\,;\,
             z^{-1}\Bigr) 
\nn\\ 
 &\qquad \Bigl[a_1-a_2\not\in \mathbb{Z},\,
               a_1-a_3\not\in \mathbb{Z},\, 
               a_2-a_3\not\in \mathbb{Z},\, 
      \mbox{ and }
               z\not\in (0,1) \Bigr] \,,
\\ 
 &\, _3F_2\Bigl(
          \afrac{a_1,\,a_2,\,0}{b_1,\,b_2};\,z
         \Bigr) = 1 \,,
\end{align}
from which one can show
\begin{align} 
 &\, _3\widehat{F}_2\Biggl(
          \afrac{\frac{d-1}{2},\,
                              \frac{d-1}{2}\pm\nu,\,
                 \frac{\frac{d-1}{2}\pm\nu+\ii\Delta E}{2}}
                             {1\pm\nu,\,
                              \frac{\frac{d+3}{2}\pm\nu+\ii\Delta E}{2}};\,z
         \Biggr) 
\nn\\ 
 &= (-z)^{-\frac{1}{2} \bigl(\frac{d-1}{2} \pm \nu +\ii\Delta E\bigr)}\, 
             \sin\Bigl(\pi\,\frac{\frac{d-1}{2}\mp \nu +\ii\Delta E}{2} \Bigr)
\nn\\ 
 &\qquad\times \frac{
         \Gamma\bigl(\frac{\frac{d-1}{2} +\nu +\ii\Delta E}{2}\bigr)\, 
         \Gamma\bigl(\frac{\frac{d-1}{2} -\nu +\ii\Delta E}{2}\bigr)\, 
         \Gamma\bigl(\frac{\frac{d-1}{2} +\nu -\ii\Delta E}{2}\bigr)\, 
         \Gamma\bigl(\frac{\frac{d-1}{2} -\nu -\ii\Delta E}{2} \bigr)}{\pi} 
\nn\\ 
 &\quad +(-z)^{-\frac{d-1}{2}}\,
   \frac{\sin\bigl(\pi\, \frac{d-1 \mp 2\nu}{2} \bigr)}
        {\sin(\mp\pi\nu)}
   \, _3\widehat{F}_2\Biggl(
          \afrac{\frac{d-1}{2} ,\,
                              \frac{d-1}{2}\mp \nu ,\,
                              \frac{\frac{d-1}{2}\mp \nu-\ii\Delta E}{2}}
                             {1\mp\nu,\,
                              \frac{\frac{d+3}{2}\mp \nu-\ii\Delta E}{2}}\,;\,
                             z^{-1} \Biggr) 
\nn\\ 
 &\quad +(-z)^{-\frac{d-1\pm 2\nu}{2}}\,
   \frac{\sin\bigl(\pi\,\frac{d-1}{2}\bigr)}
        {\sin(\pm\pi\nu)}
   \, _3\widehat{F}_2\Biggl(
          \afrac{\frac{d-1}{2} ,\,
                              \frac{d-1}{2}\pm \nu ,\,
                              \frac{\frac{d-1}{2}\pm \nu-\ii\Delta E}{2}}
                             {1\pm\nu,\,
                              \frac{\frac{d+3}{2}\pm \nu-\ii\Delta E}{2}}\,;\,
                             z^{-1} \Biggr) \,.
\end{align}
Using this equality, the following formula can be shown to hold:
\begin{align} 
 &-\frac{e^{-\ii\pi\,(\frac{d-1}{2}+\nu)}}
       {\sin(\pi\,\nu)}\,
  \, _3\widehat{F}_2\Biggl(
          \afrac{\frac{d-1}{2},\,
                              \frac{d-1}{2}+\nu,\,
                              \frac{\frac{d-1}{2}+\nu+\ii\Delta E}{2}}
                             {1+\nu,\,
                              \frac{\frac{d+3}{2}+\nu+\ii\Delta E}{2}};\,e^{\ii0} 
         \Biggr) 
\nn\\ 
 &-\frac{e^{-\ii\pi\,(\frac{d-1}{2}-\nu)}}
        {\sin(-\pi\,\nu)}\,
  \, _3\widehat{F}_2\Biggl(
          \afrac{\frac{d-1}{2},\,
                              \frac{d-1}{2}-\nu,\,
                              \frac{\frac{d-1}{2}-\nu+\ii\Delta E}{2}}
                             {1-\nu,\,
                              \frac{\frac{d+3}{2}-\nu+\ii\Delta E}{2}};\,e^{\ii0} 
         \Biggr) 
\nn\\ 
 &= e^{-\pi \Delta E}\,
   \frac{
         \Gamma\bigl(\frac{\frac{d-1}{2} +\nu +\ii\Delta E}{2}\bigr)\, 
         \Gamma\bigl(\frac{\frac{d-1}{2} -\nu +\ii\Delta E}{2}\bigr)\, 
         \Gamma\bigl(\frac{\frac{d-1}{2} +\nu -\ii\Delta E}{2}\bigr)\, 
         \Gamma\bigl(\frac{\frac{d-1}{2} -\nu -\ii\Delta E}{2} \bigr)}{\pi} 
\nn\\ 
 &\quad +\frac{e^{\ii\pi\,(\frac{d-1}{2}+\nu)}}
              {\sin(\pi\nu)}\,
   \, _3\widehat{F}_2\Biggl(
          \afrac{\frac{d-1}{2} ,\,
                              \frac{d-1}{2}+\nu ,\,
                              \frac{\frac{d-1}{2}+\nu-\ii\Delta E}{2}}
                             {1+\nu,\,
                              \frac{\frac{d+3}{2}+\nu-\ii\Delta E}{2}}\,;\,
                             e^{-\ii0} \Biggr) 
\nn\\ 
 &\quad +\frac{e^{\ii\pi\,(\frac{d-1}{2}-\nu)}}
              {\sin(-\pi\nu)}\,
   \, _3\widehat{F}_2\Biggl(
          \afrac{\frac{d-1}{2} ,\,
                              \frac{d-1}{2}-\nu ,\,
                              \frac{\frac{d-1}{2}-\nu-\ii\Delta E}{2}}
                             {1-\nu,\,
                              \frac{\frac{d+3}{2}-\nu-\ii\Delta E}{2}}\,;\,
                             e^{-\ii0} \Biggr) \,.
\label{x-x_inverse}
\end{align}

\section{Derivation of Eq.~\eqref{Wightman_k_asympt}}
\label{app:asympto_Wightman-function}

We derive the asymptotic form 
\eqref{Wightman_k_asympt} 
of the Wightman function for each mode:
\begin{align}
 G_k^+(\eta,\eta';\eta_0) 
  &= \varphi_k(\eta;\eta_0)\,\varphi^*_k(\eta';\eta_0) \,.
\label{eq:app-mode_sum}
\end{align}
By introducing $z\equiv -k\,\eta_0$ and 
$\theta(z)\equiv z-(\pi/4)\,(1+2\nu)$, 
and using the asymptotic form of 
the Bessel and Neumann functions around $z=\infty$ (i.e., $\eta_0=-\infty$) 
(see 8.451 of \cite{GR}), $u_0$ has the expansion 
\begin{align} 
 u_0
 &= \ii\sqrt{\frac{2k}{\pi}}\,(-\eta_0)^{-\frac{d-1}{2}}\,\Bigl\{
             e^{-\ii\theta(z)}\,
  \Bigl[1+\ii\frac{\frac{d-2}{2}-\bigl(\nu^2-\frac{1}{4}\bigr)}{2z}
       -\frac{\bigl(\nu^2-\frac{1}{4}\bigr)\,\bigl(\nu^2-\frac{1}{4}-(d-2)-2m^2\bigr)}
             {8z^2}\Bigr] 
\nn\\ 
 &\qquad\qquad\qquad\qquad\quad
             +e^{\ii\theta(z)}\,\Bigl[-\ii\frac{d-2}{4z}
                         +\frac{\bigl(\nu^2-\frac{1}{4}\bigr)\,(d-4)-2m^2}{8z^2}\Bigr]
             +\mathcal{O}(z^{-3})\Bigr\} \,.
\label{eq:u0expansion}
\end{align}
$v_0$ can be obtained from this expression 
by replacing $\theta(z)$ with $\theta(z)-\pi/2$\,: 
\begin{align}
v_0
 &= -\sqrt{\frac{2k}{\pi}} (-\eta_0)^{-\frac{d-1}{2}} \Bigl\{
             e^{-\ii\theta(z)} 
      \Bigl[1+\ii\frac{\frac{d-2}{2}-\bigl(\nu^2-\frac{1}{4}\bigr)}{2z}
         -\frac{\bigl(\nu^2-\frac{1}{4}\bigr)\,\bigl(\nu^2-\frac{1}{4}-(d-2)-2m^2\bigr)}
               {8z^2}\Bigr] 
\nn\\ 
 &\qquad\qquad\qquad\qquad\qquad
     -e^{\ii\theta(z)}\,\Bigl[-\ii\frac{d-2}{4z}
               +\frac{\bigl(\nu^2-\frac{1}{4}\bigr)\,(d-4)-2m^2}{8z^2}\Bigr]
             +\mathcal{O}(z^{-3})\Bigr\} \,.
\end{align}
With the $u_0$ and $v_0$ given above, 
the wave function \eqref{eq:wave-function} is written as 
\begin{align} 
 \varphi_k(\eta;\eta_0)
 &= \frac{\sqrt{\pi}}{2}\,(-\eta)^{\frac{d-1}{2}}
\nn\\ 
 &\times \Bigl\{e^{-\ii\theta(z)}\,
    \Bigl[1+\ii\frac{\frac{d-2}{2}-\bigl(\nu^2-\frac{1}{4}\bigr)}{2z}
      -\frac{\bigl(\nu^2-\frac{1}{4}\bigr)\,\bigl(\nu^2-\frac{1}{4}-(d-2)\bigr)}{8z^2}\Bigr]\,
  H^{(1)}_\nu(-k\,\eta) 
\nn\\ 
 &~~
  -e^{\ii\theta(z)}\,\Bigl[-\ii\frac{d-2}{4z}
     +\frac{\bigl(\nu^2-\frac{1}{4}\bigr)\,(d-4)-2m^2}{8z^2}\Bigr]\,H^{(2)}_\nu(-k\,\eta)
  +\mathcal{O}(z^{-3})\Bigr\} \,. 
\label{wave-function_asympto}  
\end{align}
Then, by substituting this to \eqref{eq:app-mode_sum}, 
we obtain Eq.~\eqref{Wightman_k_asympt}. 

\section{Wightman function}
\label{app:Wightman-function}

\subsection{Analytic expression for the Wightman function}

The Wightman function 
for the instantaneous ground state, 
\begin{align}
 G^+(x,x';\eta_0)
 = \frac{1}{(2\pi)^{\frac{d-1}{2}}\,\abs{\bx-\bx'}^{\frac{d-3}{2}}}\,
    \int_0^\infty \!\!\!\rmd k\,k^{\frac{d-1}{2}}\,
             J_{\frac{d-3}{2}}\bigl(k\,\abs{\bx-\bx'}\bigr)\, 
    G_{k}^+(\eta,\eta';\eta_0) \,, 
\end{align}
has the expansion of the form 
\begin{align}
 G^+(x,x';\eta_0) = 
  \sum_{n=0}^\infty\sum^{2}_{a,b=1}
  c^{(n)}_{a,b}(\eta_0)\, G^{(n)}_{a,b}(x,x') \,, 
\label{eq_B:G+expansion}
\end{align}
where 
\begin{align} 
 &G^{(n)}_{a,b}(x,x') \equiv
 \frac{\bigl[(-\eta)\,(-\eta')\bigr]^{\frac{d-1}{2}}}
       {(2\pi)^{\frac{d-1}{2}}\,\abs{\bx-\bx'}^{\frac{d-3}{2}}}
\nn\\ 
 &\quad\times \int^{\infty}_0 \rmd k\, k^{\frac{d-1}{2}-n}\,J_{\frac{d-3}{2}}(k\abs{\bx-\bx'})\,
 H^{(a)}_\nu(-e^{\ii\sigma_a\,\varepsilon}\,k\,\eta)\,
 H^{(b)}_\nu(-e^{\ii\sigma_b\,\varepsilon}\,k\,\eta') \,, 
\label{eq:Gn-integral}
\\ 
 &c^{(0)}_{1,2}=\pi/4\,,\quad c^{(0)}_{2,1}=c^{(0)}_{1,1}=c^{(0)}_{2,2}=0\,,\quad 
  c^{(1)}_{a,b}=0\,,
\nn\\ 
 &c^{(2)}_{1,2}=c^{(2)}_{2,1}=\frac{\pi}{4}\,\Bigl(\frac{d-2}{4}\Bigr)^2\,(-\eta_0)^{-2}\,,\quad 
  c^{(2)}_{1,1}=c^{(2)}_{2,2}=0 
\label{c^n_sigmasigma'}
\end{align}
with 
\begin{align}
 \sigma_a\equiv
 \biggl\{\begin{array}{l}
     +1 \quad (a=1)\\ -1\quad (a=2)
    \end{array}\,.
\end{align}
Here, an infinitesimal positive constant $\varepsilon$ is introduced 
in order to make the $k$ integration finite.%
\footnote{
It is known that a delicate treatment is required 
in regularizing the integral to obtain the response function $\cF$  
\cite{Schlicht:2003iy,Langlois:2005nf,Satz:2006kb,Louko:2007mu,Hodgkinson:2011pc}. 
However, we expect that the current regularization is sufficient 
to obtain its derivative, $\dot{\cF}$\,, 
since it gives a de Sitter invariant form in the limit $\eta_0\to-\infty$ 
and reproduces all the known results in the literature 
as special cases. 
See comments following \eqref{eq:Fdot2} and \eqref{eq:Fdot-n=0}. 
\label{fn:regularization}
} 
Note that the $n=1$ terms will totally disappear 
as a consequence of ignoring the highly oscillating terms.

In the following, we derive an analytic expression 
for the integral \eqref{eq:Gn-integral}.  
We only take care of terms with $a\neq b$ 
since terms with $a=b$ do not contribute to the Wightman function 
at least for $n\leq 2$ [see \eqref{c^n_sigmasigma'}].

By using the identity 
$K_\nu(e^{-\ii\sigma_a\,\frac{\pi}{2}}\,z)=(\ii\sigma_a\,\pi/2)\,e^{\sigma_a\frac{\ii\pi\nu}{2}}H_\nu^{(a)}(z)$\,, 
we first obtain the relation
\begin{align} 
 &\int_0^\infty\!\! \rmd k\,k^{\frac{d-3}{2}+1-n}\,
 J_{\frac{d-3}{2}}(k\,\abs{\bx-\bx'})\,
 H_\nu^{(a)}(-k\,\eta)\,
 H_\nu^{(b)}(-k\,\eta')
\nn\\ 
 &=-\frac{4\,\sigma_a\,\sigma_b}{\pi^2}\,
 e^{-(\sigma_a+\sigma_b)\,\frac{\ii\pi\nu}{2}} \!\!
 \int_0^\infty\!\! \rmd k\,k^{\frac{d-3}{2}+1-n}\,
 J_{\frac{d-3}{2}}(k\,\abs{\bx-\bx'})\,
 K_\nu(-e^{-\ii\sigma_a\,\frac{\pi}{2}}\, k\, \eta)\,
 K_\nu(-e^{-\ii\sigma_b\,\frac{\pi}{2}}\, k\,\eta')\,.
\nn\\ 
\end{align}
Then, using the formula \cite{Bessel-Integral}
\begin{align} 
 &\int_0^\infty \rmd k\,k^{\mu+1-n}\,
 K_\nu(a\,k)\, K_\nu(b\,k)\, J_{\mu}(c\,k)  
\nn\\ 
 &=\frac{2^{\mu-1-n}}{\Gamma(\mu+1)}\,\frac{c^\mu}{b^{2\xi}} 
\nn\\ 
 &\quad \times
 \Bigl[(a/b)^{\nu}\,\Gamma(\xi)\,\Gamma(\xi+\nu)\,\Gamma(-\nu)\,
       F_4\bigl(\xi,\,\xi+\nu;\,1+\mu,\,1+\nu;\,-c^2/b^2;\,a^2/b^2\bigr)
\nn\\ 
 &\qquad  +  (a/b)^{-\nu}\,\Gamma(\xi)\,\Gamma(\xi-\nu)\,\Gamma(\nu)\,
       F_4\bigl(\xi,\,\xi-\nu;\,1+\mu,\,1-\nu;\,-c^2/b^2;\,a^2/b^2\bigr)
     \Bigr] 
\nn\\ 
 &\qquad \bigl[\xi\equiv\mu+1-(n/2)\,,\quad 
 \nu\not\in\bZZ\bigr]\,,
\end{align}
we obtain the following expression for $G^{(n)}_{a,b}(x,x')$\,:
\begin{align} 
  &G^{(n)}_{a,b}(x,x')
 =\frac{(-\sigma_a\,\sigma_b)\,e^{-(\sigma_a+\sigma_b)\,\frac{\ii\pi\nu}{2}}}
        {2^{n}\,\pi^{\frac{d+3}{2}}\,
         e^{-\ii\pi\,\sigma_b\,\frac{d-1-n}{2}}\,
         \Gamma\bigl(\frac{d-1}{2}\bigr)}\,
   \Bigl(\frac{-\eta}{-\eta'}\Bigr)^{\frac{d-1}{2}} \,(-\eta')^n
\nn\\ 
 &\quad\times
 \Bigl[\Bigl(\frac{-\eta}{-\eta'}\Bigr)^{\nu}\,
       e^{-(\sigma_a-\sigma_b)\,\frac{\ii\pi\,\nu}{2}}\,
       \Gamma\Bigl(\frac{d-1-n}{2}+\nu\Bigr)\,\Gamma\Bigl(\frac{d-1-n}{2}\Bigr)\,\Gamma(-\nu)
\nn\\ 
 &\qquad\ \times F_4\Bigl(\frac{d-1-n}{2}+\nu,
                          \frac{d-1-n}{2};\frac{d-1}{2},
                          1+\nu;
                          \frac{e^{-\ii\sigma_b\,0}\,\abs{\bx-\bx'}^2}{(-\eta')^2}\,;
                          \frac{e^{-\ii\sigma_a\,\pi}\,(-\eta)^2}{e^{-\ii\sigma_b\,\pi}\,(-\eta')^2} \Bigr)
\nn\\ 
 &\qquad\ +(\nu\to-\nu)\Bigr] \,.
\label{eq:Wightman-n}
\end{align}
Here, $F_4(a,b;c,d;x,y)$ is Appell's hypergeometric function given by
\begin{align} 
 F_4(a,b;c,d;x,y)
  &\equiv \sum_{m=0}^\infty\sum_{n=0}^\infty
  \frac{(a)_{m+n}\,(b)_{m+n}}{m!\,n!\,(c)_m\,(d)_n}\,x^m\,y^n \quad 
 \bigl[\abs{x}^{1/2}+\abs{y}^{1/2}<1\bigr]\,,
\label{eq:Appell-series}
\\ 
 (x)_n&\equiv \Gamma(x+n)/\Gamma(x)\,.
\end{align}
For $Z(x,x')\equiv \bigl[\eta^2+\eta'^2-\abs{\bx-\bx'}^2\bigr]/(2\eta\eta')<1$, 
the series in \eqref{eq:Appell-series} 
does not converge and the analytic continuation should be performed 
(see 9.185 of \cite{GR} for the integral representation of the Mellin-Barnes type).

In the limit $\abs{\bx-\bx'}\to 0$, 
one can show, by using the identity  
\begin{align}
 F_4(a,b;c,d;0,y)
 = \sum_{n=0}^\infty
  \frac{(a)_{n}\,(b)_{n}}{n!\,(d)_n}\,y^n 
 = \,_2F_1\Bigl(\afrac{a,\,b}{d};\,y\Bigr) \,,
\end{align}
that the Wightman function takes the form
\begin{align} 
 &\lim_{\abs{\bx-\bx'}\to 0}G^{(n)}_{a,b}(x,x')
\nn\\ 
 &=\frac{(\sigma_a\,\sigma_b)\,e^{-(\sigma_a+\sigma_b)\,\frac{\ii\pi\nu}{2}}}
         {2^{n}\,\pi^{\frac{d+1}{2}}\,
         e^{-\ii\pi\,\sigma_b\,\frac{d-1-n}{2}}\,
         \Gamma\bigl(\frac{d-1}{2}\bigr)}\,
   \Bigl(\frac{-\eta}{-\eta'}\Bigr)^{\frac{d-1}{2}} \,(-\eta')^n
\nn\\ 
 &\quad\times
 \Bigl[\Bigl(\frac{-\eta}{-\eta'}\Bigr)^{\nu}\,
       \frac{e^{-(\sigma_a-\sigma_b)\,\frac{\ii\pi\,\nu}{2}}}{\sin(\pi\,\nu)}\,
    \,_2\widehat{F}_1\biggl(
    \afrac{\frac{d-1-n}{2}+\nu,\frac{d-1-n}{2}}
                       {1+\nu};\,
       \frac{e^{-\ii\sigma_a\,(\pi-0)}\,(-\eta)^2}{e^{-\ii\sigma_b\,(\pi-0)}\,(-\eta')^2} \biggr)
\nn\\ 
 &\qquad\ +\Bigl(\frac{-\eta}{-\eta'}\Bigr)^{-\nu}\,
       \frac{e^{(\sigma_a-\sigma_b)\,\frac{\ii\pi\,\nu}{2}}}{\sin(-\pi\,\nu)}\,
    \,_2\widehat{F}_1\biggl(
    \afrac{\frac{d-1-n}{2}-\nu,\frac{d-1-n}{2}}
                       {1-\nu};\,
        \frac{e^{-\ii\sigma_a\,(\pi-0)}\,(-\eta)^2}{e^{-\ii\sigma_b\,(\pi-0)}\,(-\eta')^2} \biggr) \Bigr]
\nn\\ 
 &=\frac{4}{\pi}\,\frac{(-\sigma_a\,\sigma_b)\,e^{\frac{\ii\pi}{2}\,(\sigma_a+\sigma_b)\,(\frac{d-1-n}{2}-\nu)}\,
         \Gamma(\frac{d-1-n}{2})}
        {(4\pi)^{\frac{d}{2}}\,\Gamma(\frac{d-1}{2})} \,
   \bigl[(-\eta)(-\eta')\bigr]^{\frac{n}{2}} \,
\nn\\ 
 &\quad\times   \,_2\widehat{F}_1\biggl(
    \afrac{\frac{d-1-n}{2}+\nu,\frac{d-1-n}{2}-\nu}
                       {\frac{d-n}{2}};\,\frac{1-u_{a,b}}{2}\biggr)
\label{eq:Gn-x=xprime}
\end{align}
with
\begin{align}
 u_{a,b}\equiv 
   \frac{e^{-\ii\sigma_a(\pi-0)}\,(-\eta)^2 + e^{-\ii\sigma_b(\pi-0)}\,(-\eta')^2}
   {2\,e^{-\ii(\sigma_a+\sigma_b)\,(\pi-0)}\,(-\eta)\,(-\eta')} \,.
\end{align}
Here, in the last equality, we have used formulas 9.132-2 and 9.134-3 of \cite{GR}. 

We comment that \eqref{eq:Gn-x=xprime} can be obtained more directly 
by performing an integration \eqref{eq:Gn-integral} 
with setting $\abs{\bx-\bx'}= 0$ in advance:
\begin{align}
  &\lim_{\abs{\bx-\bx'}\to 0}G^{(n)}_{a,b}(x,x') =
 \frac{2\,\bigl[(-\eta)\,(-\eta')\bigr]^{\frac{d-1}{2}}}
       {(4\pi)^{\frac{d-1}{2}}\Gamma(\frac{d-1}{2})}
\int^{\infty}_0 \rmd k\, k^{d-2-n}\,
 H^{(a)}_\nu(-e^{\ii\sigma_a\,\varepsilon}\,k\,\eta)\,
 H^{(b)}_\nu(-e^{\ii\sigma_b\,\varepsilon}\,k\,\eta') \,, 
 \nn\\
&=\frac{(-\sigma_a\,\sigma_b)\,e^{-(\sigma_a+\sigma_b)\,\frac{\ii\pi\nu}{2}}\,
            \Gamma(\frac{d-1-n}{2})\,\Gamma\bigl(\frac{d-1-n}{2}-\nu\bigr)}
         {2^{n}\,\pi^{\frac{d+3}{2}}\,
         \Gamma\bigl(\frac{d-1}{2}\bigr)}\,
   (-\eta)^n\,
   e^{\ii\pi\,\sigma_a\,\frac{d-1-n}{2}}\,e^{(\sigma_a-\sigma_b)\,\frac{\ii\pi\,\nu}{2}}
\nn\\ 
 &\quad\times
    \Bigl(\frac{-\eta}{-\eta'}\Bigr)^{-\frac{d-1}{2}-\nu} \,
    \,_2\widehat{F}_1\biggl(
    \afrac{\frac{d-1-n}{2}+\nu,\frac{d-1-n}{2}}
                       {1+\nu};\,
       1-\frac{e^{-\ii\sigma_b\,(\pi-0)}\,(-\eta')^2}{e^{-\ii\sigma_a\,(\pi-0)}\,(-\eta)^2} \biggr)\,, 
\end{align}
where we have used formula 6.576-4 of \cite{GR}.
It can be found that the last expression actually coincides with the second line of \eqref{eq:Gn-x=xprime} 
by using formula 9.132-1 of \cite{GR}.

Note that the $n^{\rm th}$-order Wightman function 
$G^{(n)}_{a,b}(x,x')$ is not 
de Sitter invariant for $n>0$ 
because it has an extra factor $\bigl[(-\eta)(-\eta')\bigr]^{n/2}$ 
which multiplies a function of de Sitter invariant variable $u_{a,b}$\,.

\subsection{Behavior of the Wightman function}

\begin{figure}[htbp]
\begin{center}
\includegraphics[width=5.8cm]{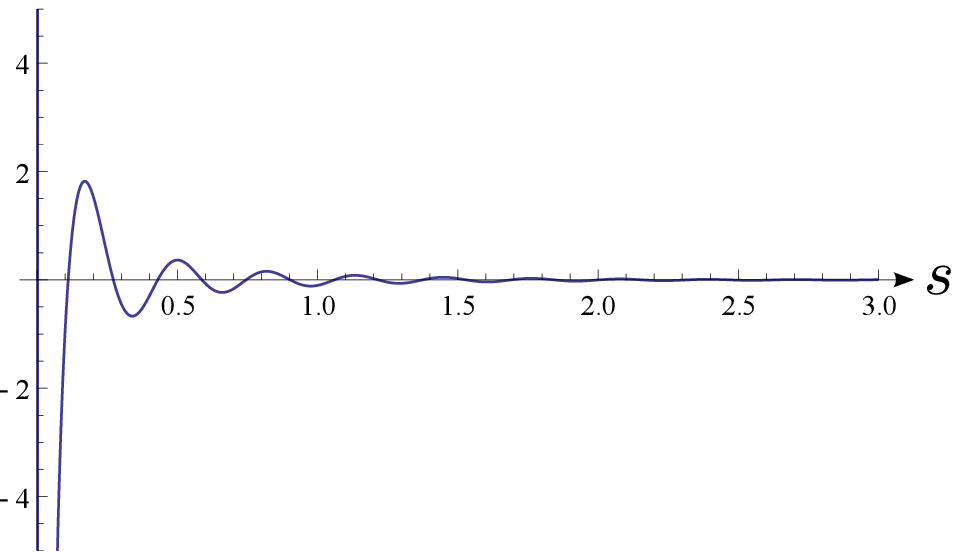}
\hspace{1cm}
\includegraphics[width=5.8cm]{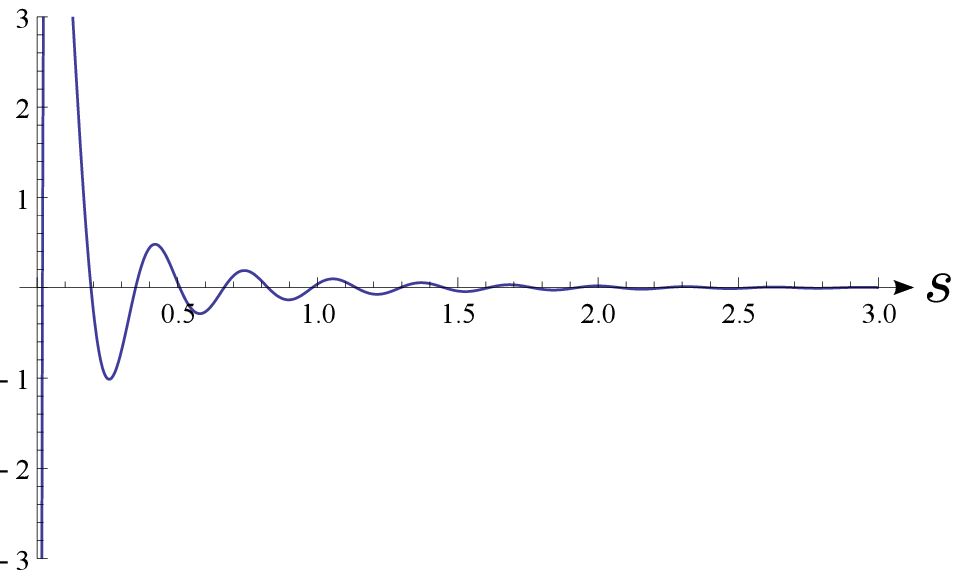}
\begin{quote}
\caption{The real part (left) and the imaginary part (right) 
of the Wightman function $G^+\bigl(x(\tau),x(\tau-s);\,\eta_0\bigr)$
for $d=4$, $\nu=20\ii$ (heavy mass case), and $\tau-\tau_0=50$. 
\label{fig:d4nu20I}
}
\end{quote}
\end{center}
\vspace{-3ex}
\end{figure}

\begin{figure}[htbp]
\begin{center}
\includegraphics[width=5.8cm]{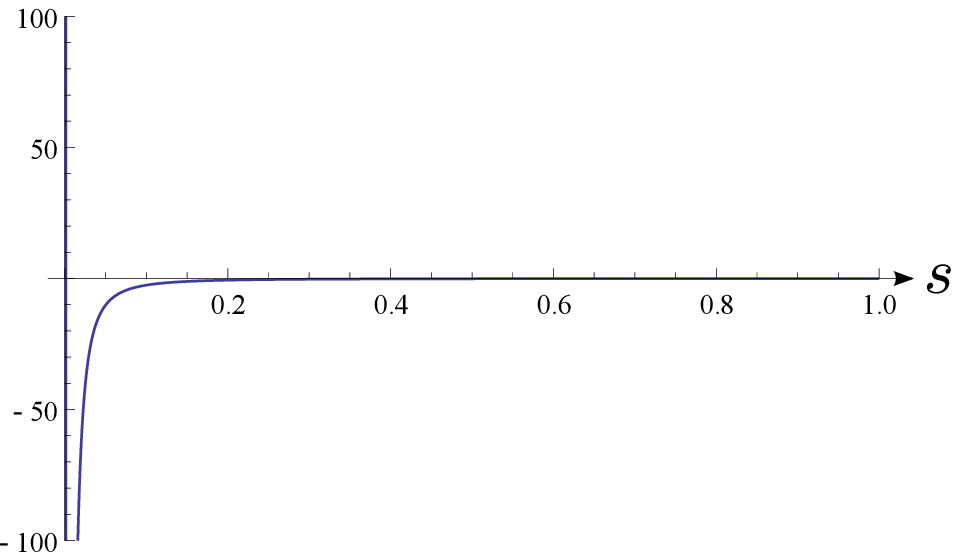}
\hspace{1cm}
\includegraphics[width=5.8cm]{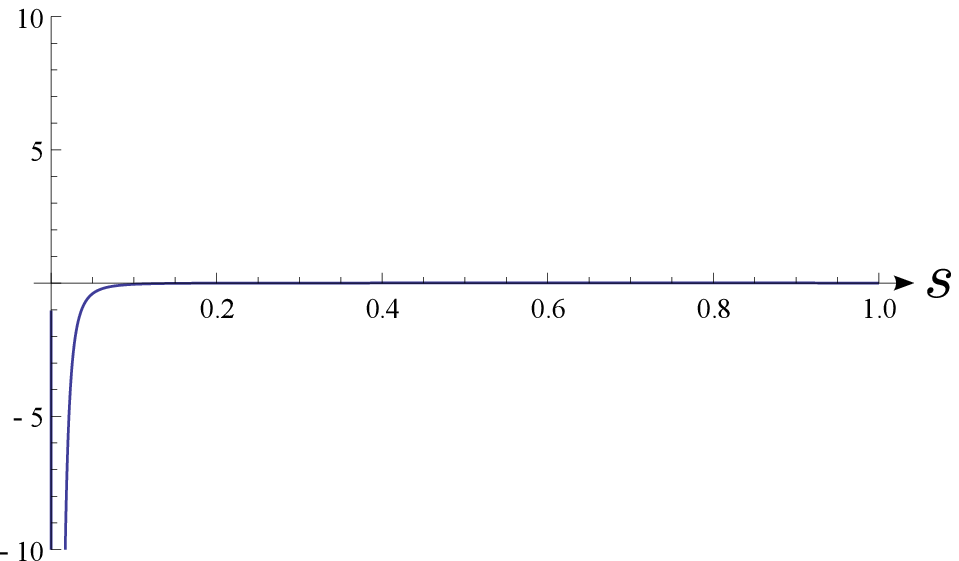}
\begin{quote}
\caption{The real part (left) and the imaginary part (right) 
of the Wightman function $G^+\bigl(x(\tau),x(\tau-s);\,\eta_0\bigr)$
for $d=4$, $\nu=0$, and $\tau-\tau_0=50$. 
\label{fig:d4nu0}
}
\end{quote}
\end{center}
\vspace{-3ex}
\end{figure}

\begin{figure}[htbp]
\begin{center}
\includegraphics[width=5.8cm]{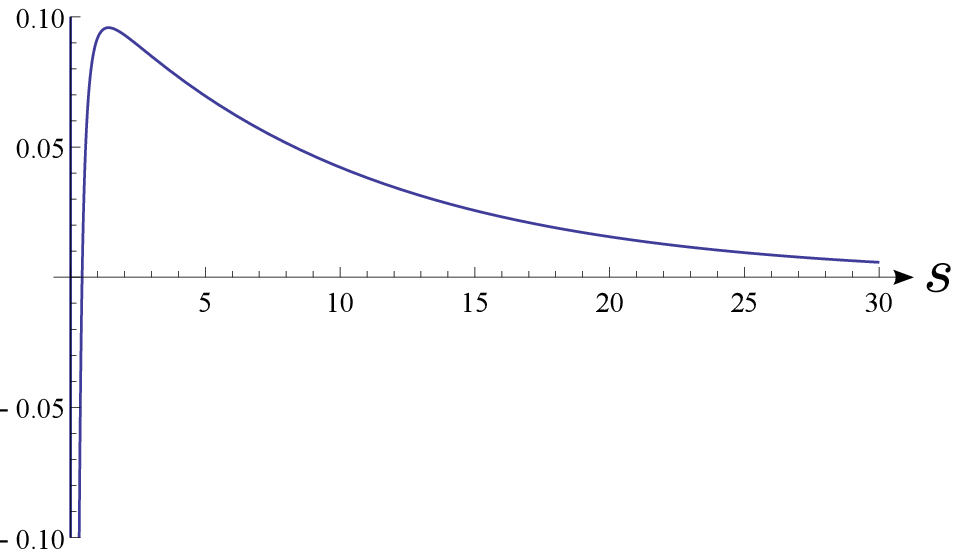}
\hspace{1cm}
\includegraphics[width=5.8cm]{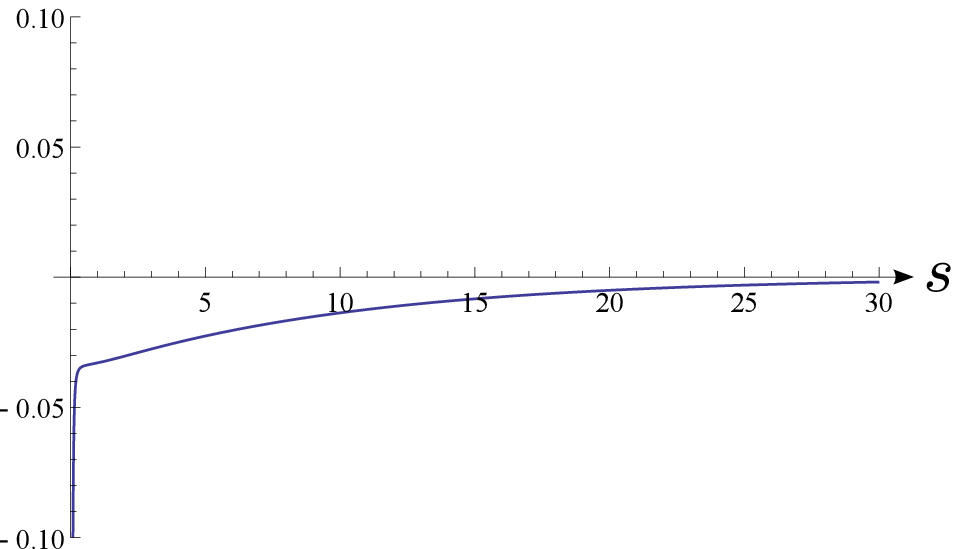}
\begin{quote}
\caption{The real part (left) and the imaginary part (right) 
of the Wightman function $G^+\bigl(x(\tau),x(\tau-s);\,\eta_0\bigr)$
for $d=4$, $\nu=1.4$ (light mass case), and $\tau-\tau_0=50$. 
\label{fig:d4nu14}
}
\end{quote}
\end{center}
\vspace{-3ex}
\end{figure}

In order to understand the behavior of the Wightman function%
\footnote{ 
The Wightman function has the following asymptotic form for large $s$\,: 
\begin{align}
 G^+\bigl(x(\tau),x(\tau-s);\,\eta_0\bigr)
 &\sim \frac{e^{-\ii\pi\,(\frac{d-1}{2}-\nu)}\,\Gamma\bigl(\frac{d-1}{2}-\nu\bigr)\,\Gamma(\nu)}{4\,\pi^{\frac{d+1}{2}}}\,
 e^{-\bigl(\frac{d-1}{2}-\nu\bigr)\,s}
\nn\\
 &\quad +\frac{e^{-\ii\pi\,(\frac{d-1}{2}+\nu)}\,\Gamma\bigl(\frac{d-1}{2}+\nu\bigr)\,\Gamma(-\nu)}{4\,\pi^{\frac{d+1}{2}}}\,
 e^{-\bigl(\frac{d-1}{2}+\nu\bigr)\,s}\,,
\nn
\end{align}
which becomes oscillatory for the heavy mass case $m>(d-1)/2$\,. 
The behavior for small $s$ is given by 
\begin{align}
 G^+\bigl(x(\tau),x(\tau-s);\,\eta_0\bigr)
 \sim \frac{e^{-\ii\pi\,\frac{d-2}{2}}\,\Gamma\bigl(\frac{d-2}{2}\bigr)}
           {4\pi^{\frac{d}{2}}}\,(s-\ii 0)^{-(d-2)}\,.
\nn
\end{align}
}
\begin{align} 
 &G^+\bigl(x(\tau),x(\tau-s);\,\eta_0\bigr) 
 \simeq  G^{(0)}_{1,2}\bigl(x(\tau),x(\tau-s)\bigr)
 \nn\\
 &\quad+ \frac{\pi}{4}\Bigl(\frac{d-2}{4}\Bigr)^2 (-\eta_0)^{-2}\, 
   [G^{(2)}_{1,2}\bigl(x(\tau),x(\tau-s)\bigr)+G^{(2)}_{2,1}\bigl(x(\tau),x(\tau-s)\bigr)]\,,
\end{align}
we give plots of the Wightman function 
for three typical cases of $d=4$ 
with $\nu=20\,\ii$ [heavy mass case; $m>(d-1)/2=3/2$\,, Fig.~\ref{fig:d4nu20I}], 
$\nu=0$ [$m=3/2$\,, Fig.~\ref{fig:d4nu0}], 
and $\nu=1.4$ [light mass case; $m<3/2$\,, Fig.~\ref{fig:d4nu14}]. 
It is clear from these examples that, 
for the cases other than the light mass case 
(i.e., for the cases described in Figs.~\ref{fig:d4nu20I} 
and \ref{fig:d4nu0}), 
the Wightman function $G^+\bigl(x(\tau),x(\tau-s);\,\eta_0\bigr)$ takes significant values 
only around the coincident point $s=0$. 
On the other hand, as can be seen in Fig.~\ref{fig:d4nu14}, 
the Wightman function has a longer-range correlation 
as the mass $m$ decreases. 
Thus, in the light mass case, we may not be able to neglect memory effects 
and the Markovian approximation used in \eqref{dense} may not be valid. 

\section{Calculation of $\dot{\cF}$}
\label{app:integration}

In this appendix, we calculate the derivative of the response function
\begin{align}
 \dot{\cF}^{(\alpha)}_{a,b}(\Delta E,\tau,\tau_1) 
 &\equiv \int_{-(\tau-\tau_1)}^0\rmd s\,
  e^{-\ii\Delta E\,s}\,G^{(\alpha)}_{a,b}(x(\tau+s),x(\tau);\eta_0)
\nn\\ 
 &\quad + \int_0^{(\tau-\tau_1)}\rmd s\,
  e^{-\ii\Delta E\,s}\,G^{(\alpha)}_{a,b}(x(\tau),x(\tau-s);\eta_0)\,.
\label{eq:Fdot-n}
\end{align}
associated with 
the Wightman function [on the trajectory \eqref{eq:geodesic}] of the form
\begin{align}
 G^{(\alpha)}_{a,b}(x,x') 
 &= \frac{2\,[(-\eta)(-\eta')]^{\frac{d-1}{2}}}{(4\pi)^{\frac{d-1}{2}}\,\Gamma(\frac{d-1}{2})}\,
    \int_0^\infty \!\!\!\rmd k\,k^{d-2-\alpha}\, 
    H^{(a)}_{\nu}(-k\,\eta) \,H^{(b)}_{\nu}(-k\,\eta')
\end{align}
Recall that $\dot{\cF}$ associated with 
the Wightman function \eqref{Wightman-decomposition} 
is given by a linear combination of $\dot{\cF}^{(\alpha)}_{a,b}$.

The first term in \eqref{eq:Fdot-n} is rewritten to
\begin{align} 
 &\int_{-(\tau-\tau_1)}^0\rmd s\,e^{-\ii\Delta E\,s}\,
    G^{(\alpha)}_{a,b}(x(\tau+s),x(\tau))
\nn\\ 
 &= \frac{2\,e^{-\alpha\,\tau}}{(4\pi)^{\frac{d-1}{2}}\,\Gamma\bigl(\frac{d-1}{2}\bigr)}\,
 \int_0^\infty \rmd \sfx \, \sfx^{\frac{d-3}{2}+\ii\Delta E}\,H^{(a)}_\nu (e^{\ii\sigma_a\varepsilon}\,\sfx)\,
 \int_{e^{-(\tau-\tau_1)}\,\sfx}^\sfx \rmd \sfy\, 
 \sfy^{\frac{d-3}{2}-\alpha-\ii\Delta E}\, 
 H^{(b)}_\nu (e^{\ii\sigma_b\,\varepsilon}\,\sfy) \,.
\label{eq:Fdot-s<0}
\end{align}
Here, we have defined new variables 
$\sfx\equiv k\,e^{-(\tau+s)}$ and $\sfy\equiv k\,e^{-\tau}$, 
and the order of integration has been changed. 
In a similar way, the second term in \eqref{eq:Fdot-n} 
is rewritten to 
\begin{align} 
 &\int_0^{\tau-\tau_1} \rmd s\,e^{-\ii\Delta E\,s}\,G^{(\alpha)}_{a,b}\bigl(x(\tau),x(\tau-s)\bigr) 
\nn\\ 
 &= \frac{2\,e^{-\alpha\,\tau}}{(4\pi)^{\frac{d-1}{2}}\,\Gamma\bigl(\frac{d-1}{2}\bigr)}\,
 \int_0^\infty \rmd \sfx \, \sfx^{\frac{d-3}{2}-\ii\Delta E}\, H^{(b)}_\nu (e^{\ii\sigma_b\,\varepsilon}\,\sfx) \,
 \int_{e^{-(\tau-\tau_1)}\,\sfx}^\sfx \rmd \sfy\, 
   \sfy^{\frac{d-3}{2}-\alpha+\ii\Delta E}\,H^{(a)}_\nu (e^{\ii\sigma_a\varepsilon}\,\sfy),
\label{eq:Fdot-s>0}
\end{align}
where we have defined $\sfx\equiv k\,e^{-(\tau-s)}$ and $\sfy\equiv k\,e^{-\tau}$, 
and again the order of integration has been changed.

The $\sfy$ integration in \eqref{eq:Fdot-s<0} and \eqref{eq:Fdot-s>0} 
can be performed by using the following formula with $p=1$ and $q=e^{-(\tau-\tau_1)}$:
\begin{align} 
 &\int_{q\,\sfx}^{p\,\sfx}\rmd \sfy\, \sfy^{\frac{d-3}{2}-\alpha\mp\ii\Delta E}\,H^{(b)}_\nu(e^{\ii\sigma_b\,\varepsilon}\,\sfy)
\nn\\ 
 &= -\frac{2\ii\sigma_b}{\pi}\,e^{-\sigma_b\frac{\ii\pi\nu}{2}} 
   \int_{q\,\sfx}^{p\,\sfx}\rmd \sfy\, \sfy^{\frac{d-3}{2}-\alpha\mp\ii\Delta E}\,K_\nu\bigl(e^{-\sigma_b\frac{\ii(\pi-2\varepsilon)}{2}}\,\sfy\bigr)
\nn\\ 
 &= -\frac{\ii\sigma_b}{\pi}\,e^{-\sigma_b\frac{\ii\pi\nu}{2}} 
   \biggl[\int_{0}^{(p\,\sfx)^2}-\int_0^{(q\,\sfx)^2}\biggr]\,
  \rmd Y\, Y^{\frac{\frac{d-5}{2}-\alpha\mp\ii\Delta E}{2}}\,
  K_\nu\Bigl(e^{-\sigma_b\frac{\ii(\pi-2\varepsilon)}{2}}\,\sqrt{Y}\Bigr)\qquad \bigl(\sfy\equiv \sqrt{Y}\bigr)
\nn\\ 
 &= -\frac{\ii\sigma_b}{\pi}\,e^{-\sigma_b\frac{\ii\pi\nu}{2}} 
  \int_0^1 \rmd \widetilde{Y}\, 
  \Bigl[(p\,\sfx)^{\frac{d-1}{2}-\alpha\mp\ii\Delta E}\,\widetilde{Y}^{\frac{\frac{d-5}{2}-\alpha\mp\ii\Delta E}{2}}\,
         K_\nu\Bigl(e^{-\sigma_b\frac{\ii(\pi-2\varepsilon)}{2}}\,a\,\sfx\,\sqrt{\widetilde{Y}}\Bigr)
        - \bigl(p\to q\bigr)\Bigr]
\nn\\ 
 &= -\frac{2^{\nu-1}\,\ii\sigma_b}{\pi}\, 
    \Biggl[(p\,\sfx)^{\frac{d-1}{2}-\nu-\alpha\mp\ii\Delta E}\,
 G^{2,1}_{1,3}\Biggl(\frac{-e^{2\ii\sigma_b\,\varepsilon}\,(a\,\sfx)^2}{4}\,\Bigg\vert
     \afrac{\frac{\nu-\frac{d-5}{2}+\alpha\pm\ii\Delta E}{2}}
                        {\nu,\,
                         0,\,
                         \frac{\nu-\frac{d-1}{2}+\alpha\pm\ii\Delta E}{2}}\Biggr)
 - (p\to q)\Biggr]\,.\nn\\ 
 \label{eq:y-H_G1321}
\end{align}
In the last equality, we have used the following formula 
(see 6.592-2 of \cite{GR}):
\begin{align} 
 \int_0^1 \rmd x\, 
  x^{\lambda}\, K_\nu(a\,\sqrt{x})
 &= 2^{\nu-1} \,a^{-\nu}\,
         G^{2,1}_{1,3}\Bigl(\frac{a^2}{4}\,\Big\vert
     \afrac{\frac{\nu}{2}-\lambda}
                        {\nu,\,
                         0,\,
                         \frac{\nu}{2}-\lambda-1}\Bigr) 
\nn\\ 
 &\qquad \Bigl[{\rm Re}\,\lambda>-1 +\frac{\bigl\vert{\rm Re}\,\nu \bigr\vert}{2}\Bigr]\,.
\end{align}
If $q\neq 0$, the condition 
${\rm Re}\,\lambda>-1 +(\bigl\vert{\rm Re}\,\nu \bigr\vert/2)$ 
$[\,\Leftrightarrow (d-1)/2> \bigl\vert{\rm Re}\,\nu \bigr\vert + \alpha\,]$
is not necessary.

In order to perform the $\sfx$ integral in \eqref{eq:Fdot-s<0} and \eqref{eq:Fdot-s>0}, 
we use 
\begin{align} 
 &\int_0^{\infty} \rmd x\, x^{\alpha-1}\,H^{(a)}_{\nu}(x)\,
 G^{2,1}_{1,3}\Bigl(\frac{\omega\,x^2}{4}\,\Big\vert 
 \afrac{a_1}
                    {b_1,\, b_2,\, b_3} \Bigr) 
\nn\\ 
 &= 2^{\alpha -1}\,\Biggl[ G^{2,2}_{3,3}\biggl(\omega\,\bigg\vert 
 \afrac{-\frac{\alpha+\nu-2}{2},\,a_1,-\frac{\alpha-\nu-2}{2}}
                    {b_1,\, b_2,\, b_3} \biggr) 
 +\ii\sigma_a\, G^{2,3}_{4,4}\biggl(\omega\,\bigg\vert 
 \afrac{-\frac{\alpha+\nu-2}{2},\,
                     -\frac{\alpha-\nu-2}{2},\,
                     a_1,\,
                     -\frac{\alpha-\nu-3}{2}}
                    {b_1,\, b_2,\, b_3,\,-\frac{\alpha-\nu-3}{2}} \biggr)\Biggr]\,,
\nn\\  &
\label{eq:y-H-G_G}
\end{align}
which can be derived from formulas 7.821-1 and 7.821-2 of \cite{GR},
\begin{align} 
 &\int_0^{\infty}\!\! \rmd x\, x^{\alpha-1}\,J_{\nu}(x)\,
 G_{1,3}^{2,1}\Bigl(\frac{\omega\,x^2}{4}\,\Big\vert 
 \afrac{a_1}
                    {b_1,\, b_2,\, b_3} \Bigr)
 = 2^{\alpha -1}\, G_{3,3}^{2,2}\biggl(\omega\,\bigg\vert 
 \afrac{-\frac{\alpha+\nu-2}{2},\,a_1,-\frac{\alpha-\nu-2}{2}}
                    {b_1,\, b_2,\, b_3} \biggr)\,,
\\ 
 &\int_0^{\infty}\!\! \rmd x\, x^{\alpha-1}\,N_{\nu}(x)\,
 G_{1,3}^{2,1}\Bigl(\frac{\omega\,x^2}{4}\,\Big\vert 
 \afrac{a_1}
                    {b_1,\, b_2,\, b_3} \Bigr)
 = 2^{\alpha -1}\, G_{4,4}^{2,3}\biggl(\omega\,\bigg\vert 
 \afrac{-\frac{\alpha+\nu-2}{2},\,
                     -\frac{\alpha-\nu-2}{2},\,
                     a_1,\,
                     -\frac{\alpha-\nu-3}{2}}
                    {b_1,\, b_2,\, b_3,\,-\frac{\alpha-\nu-3}{2}} \biggr)\,.
\end{align}
We thus have
\begin{align}
 &\frac{2\,e^{-\alpha\,\tau}}{(4\pi)^{\frac{d-1}{2}}\,\Gamma\bigl(\frac{d-1}{2}\bigr)}\,
 \int_0^{\infty}\rmd \sfx\, \sfx^{\frac{d-3}{2}\pm\ii\Delta E} 
      H_{\nu}^{(a)}(e^{\ii\sigma_a\varepsilon}\,\sfx)\,
 \int_{e^{-(\tau-\tau_1)}\,\sfx}^\sfx\rmd \sfy\, \sfy^{\frac{d-3}{2}-\alpha\mp\ii\Delta E}\,
      H^{(b)}_\nu(e^{\ii\sigma_b\,\varepsilon}\,\sfy) 
\nn\\ 
 &= -\frac{\ii\sigma_b\,e^{-\alpha\,\tau}}
          {2^{\alpha+1}\,\pi^{\frac{d+1}{2}}\,\Gamma\bigl(\frac{d-1}{2}\bigr)}
\nn\\ 
 &\times  \Biggl[ G^{2,2}_{3,3}\Biggl(-e^{-2 \ii\,(\sigma_a-\sigma_b)\,\varepsilon} \,\Big\vert 
 \afrac{-\frac{d-3-\alpha}{2},\,
                     \frac{\nu-\frac{d-5}{2}+\alpha\pm\ii\Delta E}{2},\,
                     -\frac{d-3-\alpha}{2}+\nu}
                    {\nu,\,
                     0,\,
                     \frac{\nu-\frac{d-1}{2}+\alpha\pm\ii\Delta E}{2}}\Biggr)
\nn\\ 
 &  +\ii \sigma_a\,G^{2,3}_{4,4}\Biggl(-e^{-2 \ii\,(\sigma_a-\sigma_b)\,\varepsilon} \,\Big\vert 
 \afrac{-\frac{d-3-\alpha}{2},\,
                     -\frac{d-3-\alpha}{2}+\nu,\,
                     \frac{\nu-\frac{d-5}{2}+\alpha\pm\ii\Delta E}{2},\,
                     -\frac{d-4-\alpha}{2}+\nu}
                    {\nu,\,
                     0,\,
                     \frac{\nu-\frac{d-1}{2}+\alpha\pm\ii\Delta E}{2},\,
                     -\frac{d-4-\alpha}{2}+\nu}\Biggr)
\nn\\ 
 & -e^{-\bigl(\frac{d-1}{2}-\nu-\alpha\mp\ii\Delta E\bigr)\,(\tau-\tau_1)}\,
   \Biggl\{G^{2,2}_{3,3}\Biggl(-e^{-2 \ii\,(\sigma_a-\sigma_b)\,\varepsilon} e^{-2(\tau-\tau_1)} \,\Big\vert 
 \afrac{-\frac{d-3-\alpha}{2},\,
                     \frac{\nu-\frac{d-5}{2}+\alpha\pm\ii\Delta E}{2},\,
                     -\frac{d-3-\alpha}{2}+\nu}
                    {\nu,\,
                     0,\,
                    \frac{\nu-\frac{d-1}{2}+\alpha\pm\ii\Delta E}{2}}\Biggr)
\nn\\ 
 & +\ii \sigma_a\,G^{2,3}_{4,4}\Biggl(-e^{-2 \ii\,(\sigma_a-\sigma_b)\,\varepsilon}
 e^{-2(\tau-\tau_1)}\,\Big\vert
 \afrac{-\frac{d-3-\alpha}{2},\,
                     -\frac{d-3-\alpha}{2}+\nu ,\,
                     \frac{\nu-\frac{d-5}{2}+\alpha\pm\ii\Delta E}{2},\,
                     -\frac{d-4-\alpha}{2}+\nu}
                    {\nu,\,
                     0,\,
                     \frac{\nu-\frac{d-1}{2}+\alpha\pm\ii\Delta E}{2},\,
                    -\frac{d-4-\alpha}{2}+\nu}\Biggr)
    \Biggr\}\Biggr] \,.
\nn\\  &
\label{eq:intHH-G}
\end{align}
Combining \eqref{eq:Fdot-n}, \eqref{eq:Fdot-s<0}, \eqref{eq:Fdot-s>0}, \eqref{eq:intHH-G} 
and \eqref{eq:G+iG-3F2}, 
we finally obtain the following expression 
for $\dot{\cF}^{(\alpha)}_{a,b}(\Delta E,\tau,\tau_1)$\,:
\begin{align}
 \dot{\cF}^{(\alpha)}_{a,b}(\Delta E,\tau,\tau_1)
 &= b^{(\alpha,0)}_{a,b}\,e^{-\alpha\,\tau}
   +b^{(\alpha,1)}_{a,b}\,e^{-(\frac{d-1}{2}+\nu -\ii\Delta E)\,\tau}
   +b^{(\alpha,2)}_{a,b}\,e^{-(\frac{d-1}{2}-\nu -\ii\Delta E)\,\tau}
\nn\\
 &\quad
   +b^{(\alpha,3)}_{a,b}\,e^{-(\frac{d-1}{2}+\nu +\ii\Delta E)\,\tau}
   +b^{(\alpha,4)}_{a,b}\,e^{-(\frac{d-1}{2}-\nu +\ii\Delta E)\,\tau}\,,
\label{eq:dot-F-general}
\end{align}
where
\begin{align}
 &b^{(\alpha,0)}_{a,b} 
 \equiv \bigl(-e^{-(\sigma_a-\sigma_b)\,\ii 0}\bigr)^\nu\,
         {\bf F}^{(\alpha)}_{a,b}(\nu,\Delta E\,;\,1)
        +e^{-\sigma_a\ii\pi\nu}\,{\bf F}^{(\alpha)}_{a,b}(-\nu,\Delta E\,;\,1)
\nn\\
 &\qquad +\bigl(-e^{(\sigma_a-\sigma_b)\,\ii 0}\bigr)^\nu\,
         {\bf F}^{(\alpha)}_{b,a}(\nu,-\Delta E\,;\,1)
        +e^{-\sigma_b\ii\pi\nu}\,{\bf F}^{(\alpha)}_{b,a}(-\nu,-\Delta E\,;\,1)
\\
 &b^{(\alpha,1)}_{a,b} 
  \equiv
 -\bigl(-e^{-(\sigma_a-\sigma_b)\,\ii 0}\bigr)^\nu\,
         e^{(\frac{d-1}{2}+\nu-\alpha -\ii\Delta E)\,\tau_1}\,
         {\bf F}^{(\alpha)}_{a,b}\bigl(\nu,\Delta E\,;\,e^{-2(\tau-\tau_1)}\bigr)\,,
\\
 &b^{(\alpha,2)}_{a,b} 
  \equiv
 -e^{-\sigma_a\ii\pi\nu}\,
         e^{(\frac{d-1}{2}-\nu-\alpha -\ii\Delta E)\,\tau_1}\,
         {\bf F}^{(\alpha)}_{a,b}\bigl(-\nu,\Delta E\,;\,e^{-2(\tau-\tau_1)}\bigr)\,,
\\
 &b^{(\alpha,3)}_{a,b} 
  \equiv
 -\bigl(-e^{(\sigma_a-\sigma_b)\,\ii 0}\bigr)^\nu\,
         e^{(\frac{d-1}{2}+\nu-\alpha +\ii\Delta E)\,\tau_1}\,
         {\bf F}^{(\alpha)}_{b,a}\bigl(\nu,-\Delta E\,;\,e^{-2(\tau-\tau_1)}\bigr)\,,
\\
 &b^{(\alpha,4)}_{a,b} 
  \equiv
 -e^{-b\ii\pi\nu}\,
         e^{(\frac{d-1}{2}-\nu-\alpha +\ii\Delta E)\,\tau_1}\,
         {\bf F}^{(\alpha)}_{b,a}\bigl(-\nu,-\Delta E\,;\,e^{-2(\tau-\tau_1)}\bigr)\,,
\\
 &{\bf F}^{(\alpha)}_{a,b}\bigl(\nu,\Delta E\,;\,x\bigr)
  \equiv \frac{e^{\sigma_a\ii\pi\frac{d-\alpha}{2}}e^{-\sigma_b\ii\frac{\pi}{2}}}{2^{\alpha+1}\,\pi^{\frac{d+1}{2}}\,\Gamma(\frac{d-1}{2})\,\sin(\pi\nu)}\,
\nn\\
 &\qquad\qquad\qquad \times \, _3\widehat{F}_2\Biggl(
         \afrac{\frac{d-1-\alpha}{2},\,
                             \frac{d-1-\alpha}{2}+\nu,\,
                             \frac{\frac{d-1}{2}+\nu-\alpha - \ii\Delta E}{2}}
                            {1+\nu,\,
                             \frac{\frac{d+3}{2}+\nu-\alpha - \ii\Delta E}{2}};\,
                             e^{-\ii(\sigma_a-\sigma_b)\,0}\,x\Biggr)\,.
\end{align}
Since $b^{(\alpha,0)}_{a,b}$ is independent of $\tau$, 
and $b^{(\alpha,i)}_{a,b}$ ($i=1,2,3,4$) 
become independent of $\tau$ in the limit $\tau\to \infty$, 
we find $\dot{\cF}^{(\alpha)}_{a,b}$ to take the following asymptotic form 
at later times:
\begin{align}
 \dot{\cF}^{(\alpha)}_{a,b} 
 \sim {\rm const.}\,e^{-\alpha\,\tau} 
  + {\rm const.}\,e^{-(\frac{d-1}{2}\pm\nu\pm\ii\Delta E)\,(\tau-\tau_1)} \,.
\end{align}
Then, for the Wightman function $\Delta G_{k}^+(\eta,\eta')$ 
given in \eqref{eq:DeltaG+}, 
the derivative of the response function becomes
\begin{align}
 \Delta\dot{\cF}(\Delta E;\tau,\tau_1) 
 &\sim \sum_{a,b=1}^2 {\rm const.} \, \dot{\cF}^{(\alpha_{ab})}_{a,b} 
\nn\\
 &\sim \sum_{a,b=1}^2\bigl[\,{\rm const.}\,e^{-\alpha_{ab}\,\tau} 
  + {\rm const.}\, e^{-(\frac{d-1}{2}\pm\nu\pm\ii\Delta E)\,(\tau-\tau_1)}\,\bigr] 
\nn\\
 &\sim {\rm const.}\,e^{-\alpha\,\tau} 
  + {\rm const.}\, e^{-(\frac{d-1}{2}\pm\nu\pm\ii\Delta E)\,(\tau-\tau_1)} 
\label{eq:DF-asympt}
\end{align}
with $\alpha\equiv \underset{a,b}{\rm min}(\alpha_{ab})$\,.

The derivative of the response function for the Bunch-Davies vacuum, 
$\dot{\cF}_{\rm BD}$\,, 
can also be calculated 
from \eqref{eq:dot-F-general} by setting $a=1$, $b=2$ and $\alpha=0$\,: 
\begin{align} 
 &\dot{\cF}_{\rm BD}(\Delta E,\tau,\tau_1)
  = \frac{\pi}{4}\,\dot{\cF}^{(0)}_{1,2}(\Delta E,\tau,\tau_1)
\nn\\
 &= \dot\cF^{\rm eq}(\Delta E) 
   +\frac{\pi}{4}\,b^{(0,1)}_{1,2}\,e^{-(\frac{d-1}{2}+\nu -\ii\Delta E)\,\tau}
   +\frac{\pi}{4}\,b^{(0,2)}_{1,2}\,e^{-(\frac{d-1}{2}-\nu -\ii\Delta E)\,\tau}
\nn\\
 &\quad
   +\frac{\pi}{4}\,b^{(0,3)}_{1,2}\,e^{-(\frac{d-1}{2}+\nu +\ii\Delta E)\,\tau}
   +\frac{\pi}{4}\,b^{(0,4)}_{1,2}\,e^{-(\frac{d-1}{2}-\nu +\ii\Delta E)\,\tau} \,,
\label{eq:BD-full}
\end{align} 
where 
\begin{align} 
 \dot\cF^{\rm eq}(\Delta E) 
  &\equiv  \frac{\pi}{4}\, b^{(0,0)}_{1,2}
\nn\\
 &= \frac{e^{-\pi\, \Delta E}\, 
 \Gamma\Bigl(\frac{\frac{d-1}{2}+\nu +\ii\Delta E}{2}\Bigr)\,
 \Gamma\Bigl(\frac{\frac{d-1}{2}-\nu +\ii\Delta E}{2}\Bigr)\,
 \Gamma\Bigl(\frac{\frac{d-1}{2}+\nu -\ii\Delta E}{2}\Bigr)\,
 \Gamma\Bigl(\frac{\frac{d-1}{2}-\nu -\ii\Delta E}{2}\Bigr)}
         {8\,\pi^{\frac{d+1}{2}}\,\Gamma\bigl(\frac{d-1}{2}\bigr)}\,,
\label{F_eq}
\\
 b^{(0,1)}_{1,2} 
  &=
 -e^{\ii\pi\nu}e^{(\frac{d-1}{2}+\nu -\ii\Delta E)\,\tau_1}\,
         {\bf F}^{(0)}_{1,2}\bigl(\nu,\Delta E\,;\,e^{-2(\tau-\tau_1)}\bigr)\,,
\\
 b^{(0,2)}_{1,2} 
  &=
 -e^{-\ii\pi\nu}e^{(\frac{d-1}{2}-\nu -\ii\Delta E)\,\tau_1}\,
         {\bf F}^{(0)}_{1,2}\bigl(-\nu,\Delta E\,;\,e^{-2(\tau-\tau_1)}\bigr)\,,
\\
 b^{(0,3)}_{1,2} 
  &=
 -e^{-\ii\pi\nu}e^{(\frac{d-1}{2}+\nu +\ii\Delta E)\,\tau_1}\,
         {\bf F}^{(0)}_{2,1}\bigl(\nu,-\Delta E\,;\,e^{-2(\tau-\tau_1)}\bigr)\,,
\\
 b^{(0,4)}_{1,2} 
  &=
 -e^{\ii\pi\nu}\,
         e^{(\frac{d-1}{2}-\nu +\ii\Delta E)\,\tau_1}\,
         {\bf F}^{(0)}_{2,1}\bigl(-\nu,-\Delta E\,;\,e^{-2(\tau-\tau_1)}\bigr)\,.
\label{b04_12}
\end{align}
We have used \eqref{x-x_inverse} to obtain \eqref{F_eq}. 
Equations \eqref{eq:BD-full}--\eqref{b04_12} give Eq.~\eqref{eq:Fdot-n=0}, 
which shows that 
$\dot{\cF}_{\rm BD}(\Delta E,\tau,\tau_1)$ has the following asymptotic form 
for $\tau\to \infty$\,:
\begin{align} 
 \dot{\cF}_{\rm BD}(\Delta E,\tau,\tau_1)
 \sim \dot\cF^{\rm eq}(\Delta E) 
   +{\rm const.}\,e^{-(\frac{d-1}{2}\pm \nu \pm \ii\Delta E)\,\tau} \,.
\label{eq:BD-asympt}
\end{align}


\end{document}